\documentclass[%
prd,
nofootinbib,
superscriptaddress
]{revtex4}
\usepackage{amsmath}
\usepackage{amssymb}
\usepackage{braket}
\usepackage{bm}
\usepackage{aas_macros}
\usepackage{ulem}
\usepackage[dvipdfmx]{graphicx}
\input{colordvi.tex}
\usepackage{color}
\begin{document}
\title{Magnetic field spectrum at cosmological recombination revisited}
\author{Shohei Saga}
\affiliation{Department of Physics and Astrophysics, Nagoya University,
Aichi 464-8602, Japan}
\email{saga.shohei@nagoya-u.jp}

\author{Kiyotomo Ichiki}
\affiliation{Department of Physics and Astrophysics, Nagoya University,
Aichi 464-8602, Japan}
\affiliation{Kobayashi-Maskawa Institute for the Origin of Particles and the Universe, Nagoya University,
Aichi 464-8602, Japan}

\author{Keitaro Takahashi}
\affiliation{University of Kumamoto, 2-39-1, Kurokami, Kumamoto 860-8555, Japan}

\author{Naoshi Sugiyama}
\affiliation{Department of Physics and Astrophysics, Nagoya University,
Aichi 464-8602, Japan}
\affiliation{Kobayashi-Maskawa Institute for the Origin of Particles and the Universe, Nagoya University,
Aichi 464-8602, Japan}
\affiliation{Kavli Institute for the Physics and Mathematics of the Universe (Kavli IPMU), The University of Tokyo,
Chiba 277-8582, Japan}
\begin{abstract}
If vector type perturbations are present in the primordial plasma before
recombination, the generation of magnetic fields is known to be
inevitable through the Harrison mechanism. In the context of the
standard cosmological perturbation theory, nonlinear couplings of
first-order scalar perturbations create second-order vector
perturbations, which generate magnetic fields.
Here we reinvestigate the generation of magnetic fields at second-order
in cosmological perturbations on the basis of our previous study, and extend it by newly taking into account the time
evolution of purely second-order vector perturbations with a newly developed
second-order Boltzmann code.
We confirm that the amplitude of magnetic fields from the product-terms
of the first-order scalar modes is consistent with the result in our
previous study. However, we find, both numerically and analytically,
that the magnetic fields from the purely second-order vector
perturbations partially cancel out the magnetic fields from one of the
product-terms of the first-order scalar modes, in the tight coupling
regime in the radiation dominated era. Therefore, the amplitude of
the magnetic fields on small scales, $k \gtrsim 10~h{\rm Mpc}^{-1}$, is
smaller than the previous estimates. The amplitude of the generated
magnetic fields at cosmological recombination is about $B_{\rm rec}
=5.0\times 10^{-24}~{\rm Gauss}$ on $k = 5.0 \times 10^{-1}~h{\rm
Mpc}^{-1}$.
Finally, we discuss the reason for the discrepancies that exist in
 estimates of the amplitude of magnetic fields among other authors.
\end{abstract}
\pacs{07.55.Db,98.80.-k}
\maketitle
\section{Introduction}
The presence of magnetic fields on large scales is established by current observations \cite{Widrow:2002ud,Giovannini:2003yn,Subramanian:2008tt,Ryu:2011hu,2011PhR...505....1K}.
Such cosmological magnetic fields coevolve with the Universe, e.g., astrophysical objects, cosmic microwave background radiation (CMB), large scale structure, and inflation.
Recent observations indicate that cosmological magnetic fields have the strength about micro-Gauss on Mpc scales (see, e.g., Refs.~\cite{Widrow:2002ud,Giovannini:2003yn,Subramanian:1998fn,2011PhR...505....1K,2012SSRv..166...37W,2013A&ARv..21...62D}, and references therein).
Moreover, the pair-echo method \cite{2010Sci...328...73N,2012ApJ...744L...7T,Takahashi:2013lba,Chen:2014rsa,Caprini:2015gga} determines the lower bound of strength in the intergalactic magnetic fields as $B \gtrsim O(10^{-22})~{\rm Gauss}$.
The remarkable progress of observations indicates that cosmological magnetic fields appear everywhere even in the cosmic voids.
Furthermore, the future experiments of the radio telescope such as Square Kilometer Array can survey much deeper and wider regions, and give us rich information about coevolution between cosmological magnetic fields and baryonic matters.
However, very interestingly, the origin of cosmological magnetic fields is not entirely revealed.

The key process related to evolution of magnetic fields is the dynamo mechanism \cite{1992ApJ...396..606K,1997A&A...322...98H,2012SSRv..166...37W,Brandenburg:2004jv}, which is the amplification mechanism of magnetic fields in the nonlinear magnetohydrodynamics.
In the stars, galaxies, and galaxy clusters, their nonlinear evolution can amplify seed magnetic fields, and the strength of seed fields is about $10^{-20} \sim 10^{-30}$ Gauss \cite{Davis:1999bt}.
Generally, the dynamo mechanism cannot generate magnetic fields from the absence of seed fields but amplify the seed fields, which should be set before the dynamo mechanism works.
Therefore, when we believe that the origin of cosmological magnetic fields is as a result of amplification of seed fields by the dynamo mechanism, seed fields must be created in the early stage of the universe, namely before cosmological recombination.

One of the candidates to create seed fields is the quantum fluctuations of the electromagnetic fields in the inflation era.
During the inflation era, the scale of fluctuations is extended beyond the Hubble horizon due to the nearly exponential expansion.
At first glance, it is possible to rely on this scenario to generate large-scale magnetic fields.
However, this scenario does not work since the standard Maxwell theory has the symmetry under the conformal transformation.
Under this symmetry, the vector fields, such as magnetic fields, undergo decaying only and become negligible.
In other words, when the conformal invariance is broken, inflationary magnetogenesis possibly works.
In many of the previous studies \cite{1992ApJ...391L...1R,Bamba:2003av,Martin:2007ue,2009JCAP...08..025D,2009JCAP...12..009K}, the authors have introduced interaction between the electromagnetic fields and the dilaton-like scalar field to break conformal invariance.
However, even in this case, there are other problems in the inflationary magnetogenesis, i.e., strong-coupling and backreaction problems \cite{2012JCAP...10..034F,Fujita:2014sna}.
These problems make the situation worse.
Therefore we can conclude that it is difficult to generate seed fields during the inflation era alone.
In the more recent studies \cite{Caprini:2014mja,Cheng:2014kga,Fujita:2015iga}, a new interaction between electromagnetic fields and axion-like pseudoscalar field is added in the context of dilaton-like magnetogenesis.
This interaction ends up with generation of helical magnetic fields on small scales.
The inverse cascade can transfer helical magnetic fields from smaller scales to much larger scales owing to conservation of the magnetic helicity.
We therefore expect the existence of magnetic fields on all scales in later epochs although the detailed numerical estimation is needed \cite{Christensson:2000sp,Banerjee:2004df}.

The cosmological phase transition is another possibility to generate seed fields (e.g., Refs.~\cite{2012SSRv..166...37W,2013A&ARv..21...62D}).
In general, phase transitions release the free energy and electric charges.
The released free energy is converted into the electric currents.
If these electric currents have rotational components, cosmological seed fields are induced at epochs of phase transitions.
However, the coherent length of seed fields generated in the cosmological phase transitions cannot exceed the Hubble horizon scale at that time due to the causality.
Therefore phase transitions alone are not able to explain observed large-scale magnetic fields \cite{1997PhRvL..79.1193J}.

Another category of generation mechanism is originated from astrophysical phenomena.
For example, the Biermann battery is one of the candidates of generation mechanism after recombination.
The gravitational force can be described as a gradient of the scalar potential and hence cannot generate vorticity.
However, the Biermann battery, which is nonadiabatic phenomena such as shocks, can generate vorticity and subsequently, seed fields are induced with the amplitude about $10^{-17}~{\rm Gauss}$ in protogalaxies \cite{2000ApJ...540..755D}, $10^{-17}\sim 10^{-14}~{\rm Gauss}$ in supernova remnants \cite{Hanayama:2005hd}, and $\sim 10^{-21}~{\rm Gauss}$ in galaxies \cite{Kulsrud:1996km}.
The Weibel instability, which is microscopic instability in the plasma, can amplify tiny seed fields at the epoch of structure formation \cite{Fujita:2005sh,Medvedev:2005ep}.
When the velocity distribution of plasma particles has an anisotropy in the phase space, the isotropized process of the velocity distribution releases the energy, and subsequently, the energy is converted into magnetic fields.
Accordingly, magnetic fields amplified by the Weibel instability have a quite large amplitude of about $10^{-7}~{\rm Gauss}$ \cite{Fujita:2005sh,Medvedev:2005ep}.
However, the Biermann battery and Weibel instability can only work with the existence of baryonic matters or astrophysical objects.
Therefore it is difficult to explain the origin of intergalactic magnetic fields or magnetic fields in the voids.

Yet another interesting mechanism of generating magnetic fields is the Harrison mechanism \cite{1970MNRAS.147..279H} in which magnetic fields are generated via vorticity of the primordial plasma.
In Ref.~\cite{2005PhRvL..95l1301T}, the authors have formulated the Harrison mechanism based on the cosmological perturbation theory in the primordial plasma which is a multicomponent system composed of photons, electrons, protons, dark matters, and neutrinos.
In this system, photons and electrons or protons interact with each other through the Compton scattering.
However, photons push electrons more frequently than protons because of the difference of scattering rates.
Accordingly, the charge separation takes place.
If there exist rotation-type electric fields, magnetic fields are generated.
However, in the linear perturbation theory, the Harrison mechanism does not work because there is no growing mode solution for the vector-mode perturbations which induce rotation-type electric fields.
In other words, the models including the active vector mode supplied by external sources,
i.e., free-streaming neutrinos \cite{Lewis:2004kg,2012PhRvD..85d3009I,Saga:2014zra}, cosmic defects \cite{Hollenstein:2007kg,Horiguchi:2015xsa},
and modified gravity with vector fields \cite{Saga:2013glg}, can generate magnetic fields via the Harrison mechanism.

Moreover, it turns out that even standard cosmological perturbations can generate magnetic fields if we take into account contributions from higher-order perturbations.
In fact, it is known that the second-order perturbation theory has not only the scalar mode but also the vector and tensor modes through the product of the first-order scalar perturbations.
Recently, the second-order cosmological perturbation theory is well established in the context of the CMB formalisms \cite{Hu:1993tc,Senatore:2008vi,Bartolo:2005kv,Bartolo:2006cu,Bartolo:2006fj,Pitrou:2010sn,Pitrou:2008hy,Beneke:2011kc,Beneke:2010eg,Naruko:2013aaa,Saito:2014bxa,Fidler:2014zwa}.
For example, the B-mode polarization is calculated based on the second-order perturbation theory while there is no B-mode polarization in linear scalar perturbations.

Recently, generation of magnetic fields via the second-order perturbation has been studied in detail \cite{Kobayashi:2007wd,Takahashi:2007ds,Maeda:2008dv,Gopal:2004ut,Nalson:2013jya}.
In these studies, the tight-coupling approximation is employed to estimate the amplitude of magnetic fields analytically.
Each study has shown that the amplitude of generated magnetic fields is about $10^{-30}\sim 10^{-27}~{\rm Gauss}$ at recombination on Mpc scales.
However, it is difficult to know the detail of the magnetic power spectrum since the tight-coupling approximation breaks down inside the horizon scale at recombination.
By solving perturbation equations up to the second order without employing the tight-coupling approximation, it is possible to analyze the power spectrum of magnetic fields.
In Ref.~\cite{Matarrese:2004kq}, the authors have evaluated the spectrum generated by the vorticity of charged particles, which are induced by the nonlinear coupling between the first-order density perturbations.
And they have found that resultant comoving magnetic fields have the amplitude of about $10^{-29}~{\rm Gauss}$ at recombination on Mpc scales.
Subsequently, in Refs.~\cite{Ichiki:2006cd,Ichiki:2007hu},
the authors have studied the Harrison mechanism including the anisotropic stress of photons.
They have found that the amplitude of magnetic fields has $10^{-20}~{\rm Gauss}$ at recombination on Mpc scales.
However, they ignore the purely second-order velocity difference between charged particles and photons in their analysis.
In Ref.~\cite{2011MNRAS.414.2354F}, the authors include the purely second-order effects for the first time and analyze the spectrum of magnetic fields on superhorizon scales.
In these studies, however, there are some discrepancies which have to be clarified.

In this paper we numerically solve the vector mode of cosmological
Einstein-Boltzmann equations at the second order including all the effects
relevant to the generation of magnetic fields, with a newly developed
numerical code. In addition we present analytic interpretations of the
shapes and time evolutions of the power spectrum of magnetic fields on sub- and
superhorizon scales, and make it clear what has caused the discrepancies
among the previous studies. The paper is organized as follows.
In Sec. II, we review the generation mechanism of
magnetic fields.
We decompose the evolution equation of magnetic fields up to the second order in terms of scalar and vector perturbations.
It is shown that the magnetic
fields are generated by the vector mode only. In Sec. III, we describe the
perturbed Einstein-Boltzmann system up to the second order to compute the
purely second-order perturbations, and show the tight-coupling
solutions of the system. Time evolutions of magnetic fields and spectra are numerically
evaluated and the analytic expressions are given
in Sec. IV. We devote Sec. V to our discussions
and conclusions.

Throughout this paper, we use the units in which $c=\hbar =1$ and the
metric 
signature as $(-, +, +, +)$. We obey the rule that the subscripts and
superscripts of the Greek characters and alphabets run from 0 to 3 and
from 1 to 3, respectively.
\section{Generation of magnetic fields}
In this section, we review basic equations for the generation of magnetic fields \cite{2005PhRvL..95l1301T,2006Sci...311..827I},
i.e., perturbation equations of photon, proton, and electron fluids.
While protons and electrons are conventionally treated as a single fluid, however, it is necessary to deal with proton and electron fluids separately in order to discuss the generation of magnetic fields.
Let us begin with the Euler equations. Those are given by
\begin{eqnarray}
&& m_{\rm p} n u_{\rm p}^{\mu} u_{{\rm p}i;\mu} - e n u_{\rm p}^{\mu} F_{i\mu}
 = C^{\rm pe}_{~i} + C^{{\rm p} \gamma}_{~i},
\label{eq: EOM_p2_0} \\
&& m_{\rm e} n u_{\rm e}^{\mu} u_{{\rm e}i;\mu} + e n u_{\rm e}^{\mu} F_{i\mu} =
C^{\rm ep}_{~i} + C^{{\rm e} \gamma}_{~i}~,
\label{eq: EOM_e2_0}
\end{eqnarray}
where $m_{{\rm p}({\rm e})}$ is the proton (electron) mass, $u_{{\rm p}({\rm e})}$ is the bulk
velocity of protons (electrons), $F_{\mu i}$ is the usual Maxwell
tensor. The thermal pressures of proton and electron fluids are
neglected.
The right-hand side of
Eqs.~(\ref{eq: EOM_p2_0}) and (\ref{eq: EOM_e2_0}) represent the
collision terms. The first terms in Eqs. (\ref{eq: EOM_p2_0}) and
(\ref{eq: EOM_e2_0}) are collision terms for the Coulomb scattering
between protons and electrons, which are given by \cite{Kobayashi:2007wd}
\begin{equation}
C^{\rm pe}_{i} = - C^{\rm ep}_i = -(u_{{\rm } i} - u_{{\rm e} i}) e^2 n^2 \eta_{\rm r}~,
\end{equation}
where 
\begin{equation}
\eta_{\rm r} = \frac{\pi e^2 m_{\rm e}^{1/2}}{(k_B T_{\rm e})^{3/2}} \ln \Lambda
\sim 9.4 \times 10^{-16} {\rm sec} \left(\frac{1+z}{10^5}\right)^{-3/2}
 \left(\frac{\ln\Lambda}{10}\right)~,
\end{equation}
is the resistivity of the plasma and $\ln\Lambda \sim \mathcal{O}(1)$ is the Coulomb logarithm, which is the almost constant parameter.
As is well known, this term acts as the diffusion term in the evolution equation
of magnetic field. The importance of the diffusion effect can be estimated
by the diffusion scale,
\begin{equation}
\lambda_{\rm diff}
\equiv \sqrt{\eta_{\rm r} \tau}
\sim 100 \left(\frac{\tau}{H_{0}^{-1}}\right)^{1/2} {\rm AU},
\label{eq: diffusion}
\end{equation}
above which magnetic field cannot diffuse in the time-scale $\tau$.
Here $H_{0} = 100h~{\rm km/s/Mpc}$ is the present Hubble parameter with $h$ being the normalized Hubble parameter.
Thus, at cosmological scales considered in this paper, this term can be
safely neglected.

The other terms expressed by $C^{{\rm p}({\rm e}) \gamma}_{~i}$ are the collision
terms for Compton scattering of protons (electrons) with photons.
Since photons scatter off electrons preferentially compared with protons
by a factor of $(m_{\rm e}/m_{\rm p})^2$, we can safely drop the term $C^{{\rm p} \gamma}_{~i}$
from the Euler equation of protons. This difference in collision terms
between protons and electrons ensures that small difference in velocity
between protons and electrons, that is, electric current, is indeed generated
once the Compton scattering becomes effective. 
In the next subsection, we derive the explicit form of the Compton scattering term.
\subsection{Compton collision term}
Let us now evaluate the Compton scattering term. In the limit of
completely elastic collisions between photons and electrons, this term
vanishes. Typically, in the regime of interest in this paper, very
little energy is transferred between electrons and photons in Compton
scatterings. Therefore, it is a good approximation to expand the
collision term systematically in powers of the energy transfer.

Let us demonstrate this specifically. We consider the collision process
\begin{equation}
\gamma(p^{\mu}) + e^{-}(q^{\mu}) \rightarrow \gamma(p'^{\mu}) + e^{-}(q'^{\mu}),
\end{equation}
where the quantities in the parentheses denote the particle momenta.
To calculate this process, we evaluate the collision term in the
Boltzmann equation of photons:
\begin{eqnarray}
C[f] &=&~\frac{a}{E(p)}\int{\frac{d^{3}p'}{(2\pi)^{3}2E(p')}\frac{d^{3}q}{(2\pi)^{3}2E_{\rm e}(q)}\frac{d^{3}q'}{(2\pi)^{3}2E_{\rm e}(q')}}\left| \mathcal{M}\right|^{2} \notag \\
&&\times (2\pi)^{4}\delta^{4}(q^{\mu}+p^{\mu}-q'^{\mu}-p'^{\mu})
\Bigl[ g_{\rm e}(\bm{q'})f(\bm{p'})(1+f(\bm{p}))-g_{\rm e}(\bm{q})f(\bm{p})(1+f(\bm{p'})) \Bigr] ~,
\end{eqnarray}
where $f(\bm{p})$ and $g_{\rm e}(\bm{q})$ are the distribution functions
of photons and electrons, $E_{\rm e}(q) = \sqrt{q^2+m_{\rm e}^2}$ is the energy of
an electron, and the delta functions enforce the energy and momentum
conservations.
And $\left| \mathcal{M}\right|^{2}$ is the scattering amplitude for Compton scattering.
We have dropped the Pauli blocking factor $(1-g_{\rm e})$.
The Pauli blocking factor can be always omitted safely in the epoch of
interest, because $g_{\rm e}$ is very small after electron-positron
annihilations.

Integrating over $q'$, we obtain
\begin{eqnarray}
C[f] &=&\frac{a}{p}\int{\frac{d^{3}p'}{(2\pi)^{3}2p'}\frac{d^{3}q}{(2\pi)^{3}2E_{\rm e}(q)}}\frac{2\pi}{2E_{\rm e}(|\bm{q}+\bm{p}-\bm{p'}|)}\left| \mathcal{M}\right|^{2} \notag\\
&&\times \delta \Bigl[ p-p'+E_{\rm e}(q)-E_{\rm e}(|\bm{q}+\bm{p}-\bm{p'}|) \Bigr]
\Bigl[ g_{\rm e}(\bm{q}+\bm{p}-\bm{p'})f(\bm{p'})(1+f(\bm{p}))-g_{\rm e}(\bm{q})f(\bm{p})(1+f(\bm{p'})) \Bigr] ~.
\end{eqnarray}
In the regime of our interest, energy transfer through the Compton
scattering is small and can be ignored in the first order density
perturbations. As we already discussed earlier, however, 
it is essential to take the second-order couplings in the
Compton scattering term into consideration for the generation of magnetic 
fields. Therefore, we expand the collision term up to the first order in 
powers of the energy transfer\footnote[1]{However, we shall keep up to the second-order terms
for the purpose of deriving the Einstein-Boltzmann system in section \ref{E-B system}.}, and
keep terms up to the second order in density perturbations. 

The expansion parameter is the energy transfer,
\begin{equation}
E_{\rm e}(q)-E_{\rm e}(|\bm{q}+\bm{p}-\bm{p'}|)\simeq \frac{(\bm{p'}-\bm{p})\cdot \bm{q}}{m_{\rm e}}-\frac{(\bm{p}-\bm{p'})}{2m_{\rm e}} ~,
\label{eq: E_difference}
\end{equation}
over the temperature of the universe. Employing $p \sim T$, we can
estimate the order of this expansion parameter as
${\cal O}(\frac{pq}{m_{\rm e} T}) \sim {\cal O}(\frac{q}{m_{\rm e}})$, which is small
when electrons are nonrelativistic. Note that, in the cosmological
Thomson regime, electrons in the thermal bath of photons are
nonrelativistic, $p \sim \frac{q^2}{2m_{\rm e}}$, and the energy of
photons is much smaller than the rest mass of a electron, $p \ll m_e$.
Thus, it also holds that $q \sim \sqrt{2 m_{\rm e} p} \gg p$, and the
second term in Eq.~(\ref{eq: E_difference}) is usually smaller than the
first one.

Now let us divide the collision integral into four parts, i.e., the
denominators of the Lorentz invariant volume, the scattering amplitude, the
delta function, and the distribution functions, and expand them due to
the expansion parameter defined above. First of all, the denominator
in the Lorentz invariant volume can be expanded to
\begin{eqnarray}
\frac{1}{E_{\rm e}(q)E_{\rm e}(|\bm{q}+\bm{p}-\bm{p'}|)}&=&\left( m_{\rm e}+\frac{1}{2m_{\rm e}}q^{2}\right)^{-1}\left( m_{\rm e}+\frac{1}{2m_{\rm e}}|\bm{q}+\bm{p}-\bm{p'}|\right)^{-1} \notag \\
&\approx& \frac{1}{m_{\rm e}^{2}}\left[ 1-\mathcal{E}_{(\frac{q}{m_{\rm e}})^{2}}-\mathcal{E}_{( \frac{pq}{m_{\rm e}^{2}})}-\mathcal{E}_{( \frac{p}{m_{\rm e}})^{2}} \right] ~,
\label{eq:11}
\end{eqnarray}
where
\begin{equation}
\mathcal{E}_{(\frac{q}{m_{\rm e}})^{2}} =\frac{q^{2}}{m_{\rm e}^{2}} ~,~~~
\mathcal{E}_{( \frac{pq}{m_{\rm e}^{2}})} =\frac{(\bm{p}-\bm{p'})\cdot \bm{q}}{m_{\rm e}^{2}} ~,~~~\\
\mathcal{E}_{( \frac{p}{m_{\rm e}})^{2}} =\frac{(\bm{p}-\bm{p'})^{2}}{2m_{\rm e}^{2}}~.
\end{equation}

Second, we consider the scattering amplitude.
Fortunately, it has been known that the leading term (zeroth
order term), obtained by multiplying together the first term in the
delta function and the zeroth-order distribution functions, is
zero. It means that we only have to keep up to the first order terms
when we expand the scattering amplitude and the energies, 
in order to keep the collision term up to the second order
\cite{1995ApJ...439..503D}.
The scattering amplitude for Compton scattering in the rest frame of the
electron is given by,
\begin{eqnarray}
|\mathcal{M}|^{2} &=&6\pi m_{\rm e}^{2}\sigma_{T} \left[ \frac{\tilde{p'}}{\tilde{p}}+\frac{\tilde{p}}{\tilde{p'}}-\sin^{2}\tilde{\beta} \right]~,
 \nonumber \\
 \cos\tilde{\beta} &=& \tilde{\bm{\hat{p}}}\cdot\tilde{\bm{\hat{p}}}^\prime~,
\end{eqnarray}
where $\tilde{p}$ and $\tilde{p'}$ are the energies of incident
and scattered photons, $\tilde{\bm{\hat{p}}}$ and $\tilde{\bm{\hat{p}}'}$
are the unit vectors of $\tilde{\bm{p}}$ and
$\tilde{\bm{p'}}$, respectively, denoting the directions of the
photons in this frame. The Lorentz transformation with electron's
velocity ($q/m_{\rm e}$) gives the following relations,
\begin{eqnarray}
\frac{p}{\tilde{p}}&=&\frac{\sqrt{1-(q/m_{\rm e})^{2}}}{1-\bm{p}\cdot\bm{q}/(pm_{\rm e})} ~,\\
p^{\mu}p_{\mu}&=&\tilde{p}^{\mu}\tilde{p}_{\mu}~.
\end{eqnarray}
Using these relations, we evaluate the scattering amplitude in the CMB frame
as \cite{1995PhDT..........H}
\begin{equation}
|\mathcal{M}|^{2}=6\pi m_{\rm e}^{2}\sigma_{T} \left[ \mathcal{M}_{0}+\mathcal{M}_{(\frac{q}{m_{\rm e}})}\right]~, \label{eq: 15}
\end{equation}
where
\begin{equation}
\mathcal{M}_{0}=1+\cos^{2}\beta ~,~~~
\mathcal{M}_{(\frac{q}{m_{\rm e}})}=-2\cos{\beta}(1-\cos{\beta})\left[ \frac{\bm{q}}{m_{\rm e}}\cdot \left( \bm{\hat{n}}+\bm{\hat{n}'}\right)\right] ~.
\end{equation}
Here $\bm{\hat{n}}$ and $\bm{\hat{n}'}$ are the unit vectors of $\bm{p}$ and $\bm{p'}$, respectively.

Third, we expand the delta function to 
\begin{eqnarray}
\delta \left[ p-p'+E_{\rm e}(q)-E_{\rm e}(q') \right]&\approx &\delta (p-p') +\left. \frac{\partial \delta\left[ p-p'+E{e}(q)-E_{\rm e}(q') \right]}{\partial p} \right|_{q=q'}\left( E_{\rm e}(q)-E_{\rm e}(q')\right) \notag\\
&&+\left. \frac{1}{2}\frac{\partial^{2}\delta\left[ p-p'+E_{\rm e}(q)-E_{\rm e}(q') \right]}{\partial p^{2}}\right|_{q=q'}\left( E_{\rm e}(q)-E_{\rm e}(q')\right)^{2} \notag \\
&=&\delta (p-p') +\frac{\partial \delta(p-p')}{\partial p'} \mathcal{D}_{(\frac{q}{m_{\rm e}})}+\frac{\partial \delta(p-p')}{\partial p'}\mathcal{D}_{(\frac{p}{m_{\rm e}})} +\frac{1}{2}\mathcal{D}^{2}_{(\frac{q}{m_{\rm e}})}\frac{\partial^{2}\delta (p-p')}{\partial p'^{2}} ~,
\end{eqnarray}
where
\begin{equation}
\mathcal{D}_{(\frac{q}{m_{\rm e}})}=\frac{(\bm{p}-\bm{p'})\cdot \bm{q}}{m_{\rm e}} ~,~~~
\mathcal{D}_{(\frac{p}{m_{\rm e}})}=\frac{(\bm{p}-\bm{p'})^{2}}{2m_{\rm e}}~.
\end{equation}

Finally, the distribution of the electron can be expanded to
\begin{equation}
g_{\rm e}(\bm{q}+\bm{p}-\bm{p'})\approx g_{\rm e}(\bm{q})+\frac{\partial g_{\rm e}}{\partial \bm{q}}\cdot (\bm{p}-\bm{p'})+\frac{1}{2}(p^{i}-p'^{i})\frac{\partial^{2}g_{\rm e}}{\partial q^{i}\partial q^{j}}(p^{j}-p'^{j}) .
\label{eq: dist_of_e}
\end{equation}
We assume that the electrons are kept in thermal equilibrium and in the
Boltzmann distribution:
\begin{equation}
g_{\rm e}(\bm{q})=n_{\rm e}\left( \frac{2\pi}{m_{\rm e}T_{\rm e}}\right)^{3/2}\exp{\left[ -\frac{(\bm{q}-m_{\rm e}\bm{v_{e}})^{2}}{2m_{\rm e}T_{\rm e}}\right]}~,
\end{equation}
where $v_e$ is the bulk velocity of electrons. The derivatives of
the distribution function with respect to the momentum are given as
\begin{eqnarray}
\frac{\partial g_{\rm e}}{\partial q^{i}} &=&-g_{\rm e}\frac{q_{i}-m_{\rm e}v_{{\rm e}i}}{m_{\rm e}T_{\rm e}} ~,\\
\frac{\partial^{2}g_{\rm e}}{\partial q^{i}\partial q^{j}} &=&-\frac{\partial g_{\rm e}}{\partial q^{j}}\frac{q^{i}-m_{\rm e}v^{i}_{\rm e}}{m_{\rm e}T_{\rm e}}-g_{\rm e}\frac{\delta^{ij}}{m_{\rm e}T_{\rm e}}~.
\end{eqnarray}
By substituting the above equations, Eq.~(\ref{eq: dist_of_e}) is written as
\begin{equation}
g_{\rm e}(\bm{q}+\bm{p}-\bm{p'})\approx g_{\rm e}(\bm{q})\left[ 1-\mathcal{F}_{(\frac{q}{m_{\rm e}})} +\frac{1}{2}\mathcal{F}^{2}_{(\frac{q}{m_{\rm e}})}-\mathcal{F}_{(\frac{p}{m_{\rm e}})}\right] ~,
\end{equation}
where 
\begin{equation}
\mathcal{F}_{(\frac{q}{m_{\rm e}})}=\frac{\bm{q}-m_{\rm e}\bm{v}_{\rm e}}{m_{\rm e}T_{\rm e}}\cdot (\bm{p}-\bm{p'}) ~,~~~
\mathcal{F}_{(\frac{p}{m_{\rm e}})}=\frac{1}{2}\frac{(\bm{p}-\bm{p'})^{2}}{m_{\rm e}T_{\rm e}} ~.
\end{equation}
Therefore, we have
\begin{eqnarray}
&&g_{\rm e}(\bm{q}+\bm{p}-\bm{p'})f(\bm{p'})(1+f(\bm{p}))-g_{\rm e}(\bm{q})f(\bm{p})(1+f(\bm{p'})) \notag\\
&&= g_{\rm e}(\bm{q})\left[ f(\bm{p'})-f(\bm{p})\right] -f(\bm{p'})g_{\rm e}(\bm{q})\mathcal{F}_{(\frac{q}{m_{\rm e}})}
-f(\bm{p'}) g_{\rm e}(\bm{q})\left[ \mathcal{F}_{(\frac{p}{m_{\rm e}})}-\frac{1}{2}\mathcal{F}^{2}_{(\frac{q}{m_{\rm e}})}\right] \notag \\
&&+g_{\rm e}(\bm{q})f(\bm{p})f(\bm{p'})\left[ -\mathcal{F}_{(\frac{q}{m_{\rm e}})} +\frac{1}{2}\mathcal{F}^{2}_{(\frac{q}{m_{\rm e}})}-\mathcal{F}_{(\frac{p}{m_{\rm e}})}\right] ~.
\end{eqnarray}

Combining altogether, we obtain the collision term expanded with respect
to the energy transfer as (note that this expansion is
not with respect to the density perturbations)
\begin{equation}
C[f]=\frac{3}{2}\pi^{2}\frac{a\sigma_{T}}{p}\int \frac{d^{3}p'}{(2\pi)^{3}p'}\int \frac{d^{3}q}{(2\pi)^{3}} \left[{\rm (0th~order) + (1st~order) + (2nd~order)} \right] ~, \label{eq: col term1}
\end{equation}
where
\begin{widetext}\begin{eqnarray}
\mbox{0th order term:}&& \nonumber\\
&& \mathcal{M}_{0}\delta (p-p')g_{\rm e}(\bm{q})\left[ f(\bm{p'})-f(\bm{p})\right] ,\\
\mbox{1st order terms:}&& \nonumber\\
&& 
\mathcal{M}_{0}g_{\rm e}(\bm{q})\left[ -\delta (p-p')f(\bm{p'})\mathcal{F}_{(\frac{q}{m_{\rm e}})}+\frac{\partial \delta (p-p')}{\partial p'}\left[ f(\bm{p'})-f(\bm{p})\right] \mathcal{D}_{(\frac{q}{m_{\rm e}})} \right] \notag \\
&&+\mathcal{M}_{(\frac{q}{m_{\rm e}})}g_{\rm e}(\bm{q})\delta (p-p')\left[ f(\bm{p'})-f(\bm{p})\right] 
 ,\\
\mbox{2nd order terms:}&& \nonumber\\
 && \mathcal{M}_{0}g_{\rm e}(\bm{q})\left[ -\delta (p-p')f(\bm{p'})\left( \mathcal{F}_{(\frac{p}{m_{\rm e}})}-\frac{1}{2}\mathcal{F}^{2}_{(\frac{q}{m_{\rm e}})}\right) +\frac{1}{2}\frac{\partial^{2}\delta (p-p')}{\partial p'^{2}} \mathcal{D}^{2}_{(\frac{q}{m_{\rm e}})}\left[ f(\bm{p'})-f(\bm{p})\right]\right. \notag \\
&&+\left. \frac{\partial \delta (p-p')}{\partial p'}\mathcal{D}_{(\frac{p}{m_{\rm e}})}\left[ f(\bm{p'})-f(\bm{p})\right] -\frac{\partial \delta (p-p')}{\partial p'}\mathcal{D}_{(\frac{q}{m_{\rm e}})} f(\bm{p'})\mathcal{F}_{(\frac{q}{m_{\rm e}})} \right] \notag \\
&&+\mathcal{M}_{(\frac{q}{m_{\rm e}})}g_{\rm e}(\bm{q})\left[ -\delta (p-p')f(\bm{p'})\mathcal{F}_{(\frac{q}{m_{\rm e}})}+\frac{\partial \delta (p-p')}{\partial p'}\mathcal{D}_{(\frac{q}{m_{\rm e}})} \left[ f(\bm{p'})-f(\bm{p})\right]\right] \notag \\
&&+\mathcal{M}_{0}g_{\rm e}(\bm{q})f(\bm{p})f(\bm{p'})\left[ -\frac{\partial \delta (p-p')}{\partial p'}\mathcal{F}_{(\frac{q}{m_{\rm e}})}\mathcal{D}_{(\frac{q}{m_{\rm e}})}
+\left( \frac{1}{2}\mathcal{F}^{2}_{(\frac{q}{m_{\rm e}})}-\mathcal{F}_{(\frac{p}{m_{\rm e}})}\right)\delta (p-p')\right] \notag \\
&&-g_{\rm e}(\bm{q})f(\bm{p})f(\bm{p'})\mathcal{M}_{(\frac{q}{m_{\rm e}})}\mathcal{F}_{(\frac{q}{m_{\rm e}})}\delta (p-p') .
\end{eqnarray}\end{widetext}
From now on, we omit the second-order terms. These terms are 
not only much smaller than the first-order terms but also 
may not contribute to the Euler equation at all (see, Ref.~\cite{Bartolo:2006cu}). 
Evaluating the first moment of the
above collision term, we obtain the Compton scattering term in the
Euler equation (\ref{eq: EOM_e2}) as
\begin{eqnarray}
C^{{\rm e} \gamma}_{i}&=&-\int \frac{d^3 p}{(2\pi)^3}p_i C[f] \nonumber \\
 &=&-\frac{4 a n_{\rm e}\sigma_{T}}{3}\rho_{\gamma}\left[(v_{{\rm e} i} - v_{\gamma
 i})+\frac{3}{4}v_{{\rm e}j} \Pi_{\gamma i}{}^{j} \right] ,
\label{eq: coll}
\end{eqnarray}
where the product of the velocity of electrons and anisotropic stress of photons in Eq.~(\ref{eq: coll}) is the anisotropic part of  ``radiation drag'' in the context of the radiation hydrodynamics.
The radiation drag is originated by the electron motion in anisotropic radiation fields with absorptions and emissions.
The velocity of electrons $v_{{\rm e}i}$ obeys the Euler
equation given in the next section, and the velocity and anisotropic stress of photons, $v_{\gamma i}$ and $\Pi_{\gamma i}{}^{j}$, obey the Boltzmann equation.
We will show the explicit equations for electrons and photons in Sec. \ref{sec: Boltzmann eq}.
And we will show the anisotropic stress of photons can be written in terms of the brightness function, which is also defined in Sec.~\ref{sec: Boltzmann eq}.

Here moments of the distribution functions are given by
\begin{eqnarray}
&& \int \frac{d^{3}p}{(2\pi)^{3}} p f_{\gamma}(\bm{p}) = \rho_{\gamma}, \label{eq: moment1}\\
&& \int \frac{d^{3}p}{(2\pi)^{3}} p_{i} f_{\gamma}(\bm{p})
 = \frac{4}{3} \rho_{\gamma} v_{\gamma i}, \label{eq: moment2}\\
&& \int \frac{d^{3}q}{(2\pi)^{3}} g_{\rm e}(\bm{q})
 = n_{\rm e}, \label{eq: moment3}\\
&& \int \frac{d^{3}q}{(2\pi)^{3}} q_{i} g_{\rm e}(\bm{q})
 = \rho_{e} v_{{\rm e} i}, \label{eq: moment4}\\
&& \int \frac{d^{3}p}{(2\pi)^{3}} p^{-1} p_{i} p_{j} f_{\gamma}(\bm{p})
 = \rho_{\gamma} \Pi_{\gamma ij} + \frac{1}{3} \rho_{\gamma} \delta_{ij}, \label{eq: moment5}
\end{eqnarray}
where $\rho_{\gamma}$ and $\rho_{\rm e}(=m_{\rm e} n_{\rm e})$ are energy densities of
photons and electrons, $v^i_\gamma$ and $v^i_{\rm e}$ are their bulk three velocities
defined by $v^i \equiv u^i / u^0$, and $\Pi^{ij}_\gamma$ is anisotropic stress
of photons. 
Hereafter we use the relative velocity between photons and electrons as $\delta v_{\gamma {\rm b} i} \equiv v_{\gamma i}-v_{{\rm e} i}$.
It should be noted that the collision term (\ref{eq: coll}) was obtained
nonperturbatively with respect to density perturbations
\cite{2005PhRvL..95l1301T}. 
\subsection{Evolution equations of magnetic fields}
Now we obtain the Euler equations for protons and electrons as
\begin{eqnarray}
&& m_{\rm p} n u_{\rm p}^{\mu} u_{p i;\mu} - e n u_{\rm p}^{\mu} F_{i\mu} = 0 ~,
\label{eq: EOM_p2} \\
&& m_{\rm e} n u_{\rm e}^{\mu} u_{{\rm e} i;\mu} + e n u_{\rm e}^{\mu} F_{i\mu} = \frac{4 \sigma_{T} \rho_{\gamma} a n}{3}
\left[ \delta v_{\gamma {\rm b} i} - \frac{3}{4} v_{{\rm e}j} \Pi_{\gamma i}{}^{j} \right] ~,
\label{eq: EOM_e2}
\end{eqnarray}
where $m_{\rm p}$ is the proton mass. Here, we ignore the pressures of
proton and electron fluids. In addition, the Coulomb collision term is
neglected as explained below Eq.~(\ref{eq: diffusion}). Note that the
collision term was not evaluated in a manifestly covariant way. Here
the left-hand side in Eqs. (\ref{eq: EOM_p2}) and (\ref{eq: EOM_e2})
should be evaluated in a conformal coordinate system. We also assumed
the local charge neutrality: $n = n_{\rm e} \sim n_{\rm p}$. In the case
without electromagnetic fields ($F_{i\mu}=0$), the sum of the
equations (\ref{eq: EOM_p2}) and (\ref{eq: EOM_e2}) gives the Euler
equation for the baryons in the standard perturbation theory. On the
other hand, subtracting Eq.~(\ref{eq: EOM_p2}) multiplied by $m_{\rm e}$
from Eq.~(\ref{eq: EOM_e2}) multiplied by $m_{\rm p}$, we obtain
\begin{eqnarray}
&& - \frac{m_{\rm p}m_{\rm e}}{e}
 \left[ n u^{\mu} \left( \frac{j_{i}}{n} \right)_{;\mu}
 + j^{\mu}
 \left( \frac{m_{\rm p} - m_{\rm e}}{m_{\rm p} + m_{\rm e}} \frac{j_{i}}{en} - u_{i} \right)_{;\mu}
 \right]
 + e n ( m_{\rm p} + m_{\rm e} ) u^{\mu} F_{i\mu}
 - ( m_{\rm p} - m_{\rm e} ) j^{\mu} F_{i\mu} \nonumber \\
&& = \frac{4 m_{\rm p} \rho_{\gamma} a n \sigma_{T}}{3}
 \left[ \delta v_{\gamma {\rm b} i} - \frac{3}{4} v_{{\rm e}j}
	\Pi_{\gamma i}{}^{j} \right],
\label{eq: Ohm1}
\end{eqnarray}
where $u^{\mu}$ and $j^{\mu}$ are the center-of-mass 4-velocity of the proton and
electron fluids and the net electric current, respectively, defined as
\begin{eqnarray}
&& u^{\mu} \equiv \frac{m_{\rm p}u_{\rm p}^{\mu} + m_{\rm e}u_{\rm e}^{\mu}}{m_{\rm p} + m_{\rm e}}, \\
&& j^{\mu} \equiv e n (u_{\rm p}^{\mu} - u_{\rm e}^{\mu}).
\end{eqnarray}
Employing the Maxwell equations $F^{\mu\nu}{}_{;\nu} = j^{\mu}$, we see that
the quantities in the square bracket in the left-hand side of Eq.~(\ref{eq: Ohm1}) is
suppressed at the recombination epoch, compared to the second term,
by a factor \cite{1994MNRAS.271L..15S}
\begin{equation}
\frac{c^{2}}{L^{2} \omega_{\rm p}^{2}}
\sim 3 \times 10^{-40}
 \left( \frac{10^{3} {\rm cm}^{-3}}{n} \right)
 \left( \frac{1 {\rm Mpc}}{L} \right)^{2},
\end{equation}
where $c$ is the speed of light, $L$ is a characteristic length of the system
and $\omega_{\rm p} = \sqrt{4 \pi n e^{2}/m_{\rm e}}$ is the plasma frequency.

The third term in the left-hand side of Eq.~(\ref{eq: Ohm1}), 
i.e., $(m_{\rm p}-m_{\rm e})j^\mu F_{i\mu}$, is the Hall term
which can also be neglected because the Coulomb coupling between protons
and electrons is so tight that $|u^{i}| \gg |u_{\rm p}^{i} -
u_{\rm e}^{i}|$. Then we obtain a generalized Ohm's law: 
\begin{equation}
u^{\mu} F_{i\mu}
= \frac{4 \sigma_{T} \rho_{\gamma} a}{3 e}
 \left[ \delta v_{\gamma {\rm b} i} - \frac{3}{4} v_{{\rm e}j} \Pi_{\gamma
 i}^{\phantom{\gamma}j} \right]
\equiv C_{i}.
\end{equation}

Now we derive the evolution equation for the magnetic field, which 
can be obtained from the Bianchi identities $F_{[\mu\nu,\lambda]} =
0$, as
\begin{eqnarray}
0
&=& \frac{3}{2} \epsilon^{ijk} u^{\mu} F_{[jk,\mu]} \nonumber \\
&=& u^{\mu} {\cal B}^{i}_{~,\mu}
 - \epsilon^{ijk} \left( C_{j,k} + \frac{u^{0}_{~,j}}{u^{0}} C_{k} \right)
 - ( u^{i}_{~,j} {\cal B}^{j} - u^{j}_{~,j} {\cal B}^{i} )
 + \frac{u^{0}_{~,j}}{u^{0}} ( {\cal B}^{j} u^{i} - {\cal B}^{i} u^{j} ),
\label{eq: Bianchi}
\end{eqnarray}
where $\epsilon^{ijk}$ is the Levi-Civit\`{a} tensor and ${\cal B}^i
\equiv (a^2 B^{i}) = \epsilon^{ijk} F_{jk}/2$ is the magnetic field in
the comoving frame \cite{1998PhRvD..57.3264J}. We will now expand the
photon energy density, fluid velocities, and photon anisotropic stress
with respect to the density perturbation as
\begin{eqnarray}
&& \rho_{\gamma}(t,x_i) = \rho^{(0)}_{\gamma}(t) + \rho^{(1)}_{\gamma}(t,x_i) + \cdots, ~~~
 u^{0}(t,x_i) = a(t)^{-1} + u^{(1)0}(t,x_i) + \cdots, \nonumber \\
&& u^{i}(t,x_i) = u^{(1)i}(t,x_i) + \frac{1}{2}u^{(2)i}(t,x_i) + \cdots, ~~~
 v_{i}(t,x_i) = v^{(1)}_{i}(t,x_i) + \frac{1}{2}v^{(2)}_{i}(t,x_i) + \cdots \nonumber \\
&& \Pi_{\gamma}^{ij}(t,x_i) = \Pi^{(1)ij}_{\gamma}(t,x_i) + \cdots,
\end{eqnarray}
where the superscripts $(0), (1)$, and $(2)$ denote the order of expansion and $t$
is the cosmic time. Remembering that ${\cal B}^{i}$ is a second-order quantity, we see that
all terms involving ${\cal B}^{i}$ in Eq.~(\ref{eq: Bianchi}), other than the first term,
can be neglected. Thus we obtain
\begin{eqnarray}
\frac{d{\cal B}^{i}}{dt}
&\sim& \epsilon^{ijk} \left( C_{j,k} + \frac{u^{0}_{~,j}}{u^{0}} C_{k} \right) \nonumber \\
&=& \frac{4 \sigma_{T} \rho^{(0)}_{\gamma} a}{3 e} \epsilon^{ijk}
\Biggl[ \frac{1}{2}\delta v^{(2)}_{\gamma {\rm b} j,k} -\delta^{(1)}_{\gamma,j}\delta v^{(1)}_{\gamma {\rm b} k}
- \frac{3}{4} \left( v^{(1)}_{{\rm e}l} \Pi^{(1)l}_{\gamma j} \right)_{,k}\Biggr],
\label{eq: B_dot}
\end{eqnarray}
where we used the density contrast of photons, 
$\delta^{(1)}_{\gamma,k} \equiv \rho^{(1)}_{\gamma,k} /\rho^{(0)}_{\gamma}$. 
Further, we employed the fact that there is no vorticity in the linear order:
$\epsilon^{ijk} v^{(1)}_{j,k} = 0$. It should be noted that the velocity
of electron fluid can be approximated to the center-of-mass velocity at this order,
$v_{\rm e}^{(1)i} \sim v^{(1)i}_{\rm b}$. 
The physical meaning of this equation is that 
electrons gain (or lose) their momentum through scatterings due
to the relative velocity to photons, and the
anisotropic stress of photons. The momentum transfer from the
photons ensures the velocity difference between electrons and protons,
and thus eventually generates magnetic fields. 
We found that the contribution from the curvature perturbation is
always much smaller than that from the density contrast of photons.
Furthermore the tensor perturbation, i.e., primordial gravitational waves, is subdominant comparing with the scalar perturbation in the current observations \cite{Ade:2013zuv,Ade:2014xna,2013ApJS..208...19H}.
Therefore, we have omitted the curvature perturbation and the tensor perturbation in Eq.~(\ref{eq: B_dot}) when considering the evolution of magnetic fields.
Equation (\ref{eq: B_dot}) shows that the magnetic field cannot
be generated 
in the first order. The right-hand side of Eq.~(\ref{eq: B_dot})
contains two types of source terms, i.e., a purely second-order
term and those that consist of the products of first order
quantities. 

The first term in Eq.~(\ref{eq: B_dot}) is exactly the same as that
discussed in \cite{2005MNRAS.363..521G}. They have estimated the
amplitude of magnetic fields from these terms by considering typical
values at recombination. Here, we solve the equation numerically and obtain 
a robust prediction of the amplitude of magnetic fields in the
standard $\Lambda$CDM cosmology.
\subsection{Scalar, vector, and tensor decomposition}
We devote this subsection to rewriting the
evolution equation of magnetic fields in the context of the scalar,
vector, and tensor decomposition approach. In the standard cosmological
perturbation theory, the perturbations can be decomposed
into three modes,
i.e., scalar, vector, and tensor modes \cite{2012arXiv1210.2518S} as
\begin{eqnarray}
\omega_{i}(\bm{k}) &=& \omega_{0}(\bm{k})O^{(0)}_{i}(\bm{\hat{k}}) + \sum_{\lambda = \pm 1}\omega_{\lambda}(\bm{k})O^{(\lambda)}_{i}(\bm{\hat{k}}) ~,\\
\chi_{ij}(\bm{k}) &=& \chi_{\rm iso}(\bm{k})\delta_{ij} + \chi_{0}(\bm{k})O^{(0)}_{ij}(\bm{\hat{k}}) + \sum_{\lambda = \pm 1}\chi_{\lambda}(\bm{k})O^{(\lambda)}_{ij}(\bm{\hat{k}}) + \sum_{\sigma = \pm 2}\chi_{\sigma}(\bm{k})O^{(\sigma)}_{ij}(\bm{\hat{k}}) ~,
\end{eqnarray}
where $O^{(m)}_{i}$ and $O^{(m)}_{ij}$ are the orthogonal bases for
Fourier modes with wave number $\bm{k} \equiv k \bm{\hat{k}}$ and $m = 0$, $\pm 1$, and $\pm
2$ represent the scalar, vector, and tensor modes, respectively. 
Note that, in our notations, the normalization is different from
that adopted in Ref.~\cite{2012arXiv1210.2518S}.
These inverses are given as 
\begin{eqnarray}
&& \omega_{0}(\bm{k}) = -O^{(0)}_{i}(\bm{\hat{k}})\omega_{i}(\bm{k})~,~~~
\omega_{\lambda}(\bm{k}) = -O^{(-\lambda)}_{i}(\bm{\hat{k}})\omega_{i}(\bm{k})~, \\
&& \chi_{0}(\bm{k}) = \frac{3}{2}O^{(0)}_{ij}(\bm{\hat{k}})\chi_{ij}(\bm{k})~,~~~
\chi_{\lambda}(\bm{k}) = -2O^{(-\lambda)}_{ij}(\bm{\hat{k}})\chi_{ij}(\bm{k})~,~~~
\chi_{\sigma}(\bm{k}) = \frac{2}{3}O^{(-\sigma)}_{ij}(\bm{\hat{k}})\chi_{ij}(\bm{k})~.
\end{eqnarray}
In this notation, magnetic fields can be decomposed into the scalar and vector modes, i.e., $\mathcal{B}_{i} = \mathcal{B}_{0}O^{(0)}_{i} + \sum_{\lambda = \pm 1}\mathcal{B}_{\lambda}O^{(\lambda)}_{i}$.
When we pull out the scalar mode from Eq.~(\ref{eq: B_dot}), we find that
the right hand side of Eq.~(\ref{eq: B_dot}) vanishes, namely, 
\begin{equation}
\frac{d\mathcal{B}_{0}}{dt} = 0~.
\end{equation}
This is because magnetic fields consist of rotation of the vector
potential, or in other words, magnetic fields do not have the scalar component.
In contrast, the vector mode for Eq.~(\ref{eq: B_dot}) is given by
\begin{eqnarray}
\frac{d\mathcal{B}_{\lambda}(\bm{k})}{dt} &=&
\frac{4 \sigma_{T} \rho^{(0)}_{\gamma} a}{3 e} (\lambda k)
\Biggl[ -\frac{1}{2}\delta v^{(2)}_{\gamma {\rm b}~\lambda}(\bm{k})
+\int{\frac{d^{3}k_{1}}{(2\pi)^{3}}}\delta v^{(1)}_{\gamma {\rm b}~0}(k_{1})\delta^{(1)}_{\gamma}(k_{2})\sqrt{\frac{4\pi}{3}}Y^{*}_{1,\lambda}(\bm{\hat{k}}_{1}) \notag \\
&&-\int{\frac{d^{3}k_{1}}{(2\pi)^{3}}}\frac{5}{4}v^{(1)}_{{\rm b}~0}(k_{1})\Pi^{(1)}_{\gamma~0}(k_{2})\mathcal{Y}^{1,2}_{1,\lambda}(\bm{\hat{k}}_{1},\bm{\hat{k}}_{2})\Biggr] ~, \label{eq: vec mag}
\end{eqnarray}
where we define the function $\mathcal{Y}^{\ell_{1},\ell_{2}}_{\ell, m}(\bm{\hat{k}}_{1}, \bm{\hat{k}}_{2})$ as
\begin{equation}
\mathcal{Y}^{\ell_{1},\ell_{2}}_{\ell,m}(\bm{\hat{k}}_{1},\bm{\hat{k}}_{2}) \equiv (-1)^{m}(2\ell + 1)\sum_{m_{1},m_{2}}
\left(
\begin{array}{ccc}
\ell_{1} & \ell_{2} & \ell \\
0 & 0 & 0
\end{array}
\right)
\left(
\begin{array}{ccc}
\ell_{1} & \ell_{2} & \ell \\
m_{1} & m_{2} & -m
\end{array}
\right)
\sqrt{\frac{4\pi}{2\ell_{1}+1}}Y^{*}_{\ell_{1},m_{1}}(\bm{\hat{k}}_{1})
\sqrt{\frac{4\pi}{2\ell_{2}+1}}Y^{*}_{\ell_{2},m_{2}}(\bm{\hat{k}}_{2}) ~.
\end{equation}
where $Y_{\ell, m}(\bm{\hat{k}})$ is the spherical 
harmonics.
In the above equation, the multipoles should satisfy the condition that
$\ell_{1}+\ell_{2}+\ell={\rm even}$ because of a property of the
Wigner-3j symbol. Furthermore, the 
triangle condition, $\bm{k}_{2} = \bm{k} - \bm{k}_{1}$, is ensured
by the delta function.
Throughout this paper, we keep $\bm{k}_{2}$ in equations for simplicity of presentation even after being integrating out.
This function obeys the following relations,
\begin{equation}
\mathcal{Y}^{\ell_{1},\ell_{2}}_{\ell ,m}(\bm{\hat{k}}_{1},\bm{\hat{k}}_{2}) = \mathcal{Y}^{\ell_{2},\ell_{1}}_{\ell ,m}(\bm{\hat{k}}_{2},\bm{\hat{k}}_{1}) ~, ~~~
\mathcal{Y}^{\ell_{1},0}_{\ell ,m}(\bm{\hat{k}}_{1},\bm{\hat{k}}_{2}) = \sqrt{\frac{4\pi}{2\ell+1}}Y^{*}_{\ell,m}(\bm{\hat{k}}_{1})\delta_{\ell,\ell_{1}}~,~~~
\mathcal{Y}^{0,\ell_{2}}_{\ell ,m}(\bm{\hat{k}}_{1},\bm{\hat{k}}_{2}) = \sqrt{\frac{4\pi}{2\ell+1}}Y^{*}_{\ell,m}(\bm{\hat{k}}_{2})\delta_{\ell,\ell_{2}}~.
\end{equation}
To obtain the magnetic fields from Eq.~(\ref{eq: vec mag}), we need
to know the second-order relative velocity between baryons and photons
for the vector mode. In the next section, we show the formulation of the
second-order Einstein-Boltzmann system. 
\section{Second-order Einstein-Boltzmann system}\label{E-B system}
In this section, we derive the perturbed equations of the Einstein-Boltzmann system up to the second order.
Throughout this paper, we work in the Poisson gauge of which the line element is
\begin{equation}
ds^{2}=a^{2}(\eta)\left[ -e^{2\Psi}d\eta^{2}+2\omega_{i}d\eta dx^{i}+\left( e^{-2\Phi}\delta_{ij}+\chi_{ij}\right) dx^{i}dx^{j}\right] ~.
\end{equation}
Under the Poisson gauge, the gauge conditions $\omega^{i}{}_{,i}=\chi^{ij}{}_{, j}=0$ and the traceless condition $\chi^{i}{}_{i} = 0$ are imposed on $\omega_{i}$ and $\chi_{ij}$.
We expand the metric perturbations as 
$\Psi =
\Psi^{(1)}+\frac{1}{2}\Psi^{(2)}$, $\Phi =
\Phi^{(1)}+\frac{1}{2}\Phi^{(2)}$, $\omega_{i} =
\frac{1}{2}\omega^{(2)}_{i}$, and $\chi_{ij} =
\frac{1}{2}\chi^{(2)}_{ij}$,
where we neglect the first-order vector and tensor perturbations. The
first-order vector perturbation is neglected in the standard cosmology
because the vector perturbation has only a decaying mode in the
first-order perturbation theory. 
The first-order tensor perturbation is observationally shown to be subdominant
compared with the scalar one at first order \cite{Ade:2014xna}. 
\subsection{Boltzmann equation}\label{sec: Boltzmann eq}
From Eq.~(\ref{eq: vec mag}), full computation of magnetic fields needs
to evaluate the relative velocity between baryons and photons in the
second-order perturbation theory.
The velocity perturbations of the two fluids are described by the
Boltzmann equation. 

The Boltzmann equation with binary collisions can be written as
\begin{equation}
\frac{df}{d\lambda}(x^{\mu}, P^{\mu})=\tilde{C}\left[ f\right]~, \label{eq: Boltz1}
\end{equation}
where $\lambda$ is the affine parameter and $\tilde{C}\left[ f\right]$
is the collision term, i.e., in the case of the photon distribution
function, the Thomson interaction between photons and electrons. Note
that protons and electrons interact with each other through the Coulomb
interaction.
We can treat proton and electron fluids as a single component since these particles couple strongly by the Coulomb interaction.
For the Boltzmann equation for dark
matter or neutrinos, the collision term must vanish. 

To calculate the perturbed Boltzmann equation, it is useful to change
the coordinate system from the Poisson gauge $(x^{\mu},P^{\mu})$ to the local inertial frame $(x^{\mu}, p^{\mu})$ \cite{Senatore:2008vi}.
Since we consider the cosmological perturbations up to
the second order, the distribution function is expanded as 
\begin{equation}
f(\eta, \bm{x},p,\bm{\hat{n}})=f^{(0)}(\eta, p) + f^{(1)}(\eta, \bm{x},p,\bm{\hat{n}})+\frac{1}{2}f^{(2)}(\eta, \bm{x},p,\bm{\hat{n}})~,
\end{equation}
where $p$ and $\bm{\hat{n}}$ are the amplitude and the direction of the
photon's momentum, respectively.
The zeroth-order distribution function, $f^{(0)}(\eta, p)$, is fixed
to the Planck distribution.
It is useful to define the brightness function which is given by
\begin{equation}
\Delta^{(1,2)} (\eta, \bm{x}, \bm{\hat{n}})=\frac{\int{dp~p^{3}f^{(1,2)}}(\eta , \bm{x}, p, \bm{\hat{n}})}{\int{dp~p^{3}f^{(0)}}(\eta , p)}~,
\end{equation}
where the denominator of the right-hand side is proportional to the
mean energy density of photons. 

The angle dependence of the brightness function is expanded by the
spherical harmonics as 
\begin{equation}
\Delta^{(1,2)}(\eta, \bm{x},\bm{\hat{n}})=\sum_{\ell}\sum^{\ell}_{m=-\ell}\Delta^{(1,2)}_{\ell ,m}(\eta, \bm{x})(-i)^{\ell}\sqrt{\frac{4\pi}{2\ell +1}}Y_{\ell ,m}(\bm{\hat{n}})~.
\end{equation}
The coefficients $\Delta^{(1)}_{\ell ,m}$ are related to the density perturbation, velocity, and anisotropic stress for photons as $\Delta^{(1)}_{0,0}=\delta^{(1)}_{\gamma}$, $\Delta^{(1)}_{1,0}=4 v^{(1)}_{\gamma~0}$, and $\Delta^{(1)}_{2,0}=5\Pi^{(1)}_{\gamma~0}$, respectively, as is shown Eqs.~(\ref{eq: moment1})-(\ref{eq: moment5}).
The Boltzmann equation of photons in terms of $\Delta^{(1,2)}_{\ell, m}$
at first- and second-order is written as
\begin{equation}
\dot{\Delta}^{(1,2)}_{\ell ,m}+k\left[ \frac{c_{\ell +1, m}}{2\ell +3}\Delta^{(1,2)}_{\ell +1, m}-\frac{c_{\ell ,m}}{2\ell -1}\Delta^{(1,2)}_{\ell -1, m}\right]
=S^{(1,2)}_{\ell ,m}~,
\end{equation}
where $c_{\ell, m}\equiv \sqrt{\ell^{2}-m^{2}}$.
A dot represents a derivative with respect to the conformal time $\eta$ and here we have translated from real space to Fourier space.
The source term $S^{(2)}_{\ell ,m}$ can be expressed as
\begin{equation}
S^{(2)}_{\ell ,m}(\bm{k},\eta )=\mathcal{C}^{(2)}_{\ell, m}(\bm{k},\eta ) +\mathcal{G}^{(2)}_{\ell, m}(\bm{k},\eta ) ~, \label{scat term}
\end{equation}
Here, $\mathcal{C}^{(2)}_{\ell, m}$ is the collision term that is
proportional to $\dot{\tau}_{c}$, where $\dot{\tau}_{c}$ is the
differential optical depth which is defined by the number density of the
electron $n_{e}$, scale factor $a$, and the Thomson scattering
cross-section $\sigma_{\rm T}$ as $\dot{\tau}_{c}=-an_{e}\sigma_{\rm
T}$, and $\mathcal{G}^{(2)}_{\ell, m}$
denotes the gravitational effects, i.e., the lensing and the redshift
terms. 
The collision term $\mathcal{C}^{(2)}_{\ell, m}$ is related to Eq.~(\ref{eq: col term1}) as
\begin{equation}
\mathcal{C}^{(2)}_{\ell ,m}(\bm{k},\eta) = \int{d\Omega_{n}}(-i)^{-\ell}\sqrt{\frac{2\ell + 1}{4\pi}}Y^{*}_{\ell ,m}(\bm{\hat{n}})\frac{\int{dp~p^{3}\mathcal{C}^{(2)}}(\eta , \bm{k}, p, \bm{\hat{n}})}{\int{dp~p^{3}f^{(0)}}(\eta , p)} ~,
\end{equation}
where $\mathcal{C}^{(2)}(\eta , \bm{k}, p, \bm{\hat{n}})$ is the Fourier transformation of Eq.~(\ref{eq: col term1}),
while the gravitational effect $\mathcal{G}^{(2)}_{\ell, m}$ is coming from the left-hand side of the Boltzmann equation (\ref{eq: Boltz1})
with the same procedure as obtaining $\mathcal{C}^{(2)}(\eta , \bm{k}, p, \bm{\hat{n}})$.
In this paper, we call $\mathcal{C}_{\ell, m}$ and
$\mathcal{G}_{\ell ,m}$ the scattering term and the gravitational term,
respectively. 
The explicit forms of $\mathcal{C}^{(2)}_{\ell, m}$ and $\mathcal{G}^{(2)}_{\ell, m}$ are given in our previous study \cite{Saga:2014jca}.

The source terms of the first-order Boltzmann equation vanish
when $m\neq 0$, because we consider only the scalar mode in the first-order perturbations. 
However, for the second-order perturbations, 
not only the
scalar mode $(m=0)$, but also the vector $(m=\lambda)$ and tensor
$(m=\sigma)$ modes arise due to nonlinear couplings, where $\lambda = \pm 1$ and $\sigma = \pm 2$, respectively.

In the case of massless neutrinos, one can set $\dot{\tau}_{c}=0$
in the above equations because massless neutrinos interact with the other fluids only
through gravity. We do not write down the hierarchical equation of
neutrinos here since it is trivial. The distribution function of
neutrinos is also expanded by the spherical harmonics.
\subsection{Tight coupling solutions of the vector mode}
To solve the second-order equations derived in Sec.~\ref{sec: Boltzmann eq} numerically, we
should set up the initial condition of each perturbation
variable. Thus we first solve the equations analytically with $k\eta\ll 1$ and using the tight coupling
approximation, and find the initial condition at sufficiently early time
for our numerical calculation.

Deep in the radiation dominated era, photon and baryon fluids are tightly coupled because the opacity $\dot{\tau}_{c}$ is large \cite{Ichiki:2011ah,Maeda:2008dv,Takahashi:2007ds,Pitrou:2010ai}.
Although the photon and baryon fluids would behave as a single fluid, there is a small difference in motion between photon and baryon fluids.
For this reason, we can expand the perturbation variables using the tight-coupling parameter which is given by
\begin{equation}
\epsilon \equiv \left| \frac{k}{\dot{\tau}_{c}}\right| 
\sim 10^{-2}\left( \frac{k}{1{\rm Mpc^{-1}}}\right)\left( \frac{1+z}{10^{4}}\right)^{-2}\left( \frac{\Omega_{\rm b}h^{2}}{0.02}\right)^{-1} ~,
\end{equation}
where $\Omega_{\rm b}$ is the baryon density normalized by the critical
density, and $h$ is the normalized Hubble constant.
In what follows, we derive the tight-coupling solution up to the first
order to set the initial condition of photon and baryon fluids at second-order in cosmological perturbations and to
calculate the evolution of perturbations in a numerically stable manner.

We expand the cosmological perturbation variables using the tight-coupling parameter $\epsilon$ up to the first order as
\begin{equation}
\Delta^{({\rm CPT} = 1,2)} = \Delta^{({\rm CPT} = 1,2,{\rm TCA} = \O)} + \Delta^{({\rm CPT} = 1,2,{\rm TCA} = I)} + \cdots ~,
\end{equation}
where the Arabic number and the Roman number represent orders in the cosmological perturbation and the tight coupling expansion, respectively.
In the rest of this paper, we focus on the vector mode $m = \pm 1$.

First, the solutions at zeroth order in the tight-coupling
expansion, namely, in the tight-coupling limit, are given as
\begin{eqnarray}
\delta v^{(2,\O )}_{\gamma {\rm b}~\lambda}(\bm{k})&=&0 ~, \\
\Delta^{(2,\O )}_{2,\lambda}(\bm{k}) &=&20 \int\frac{d^{3}k_{1}}{(2\pi)^{3}} \left[ v^{(1,\O )}_{\gamma~0}(k_{1})v^{(1,\O )}_{\gamma~0}(k_{2}) \right] \mathcal{Y}^{1,1}_{2, \lambda}(\bm{\hat{k}}_{1},\bm{\hat{k}}_{2}) ~, \label{eq: TCA0 pi}\\
\Delta^{(2,\O )}_{\ell \geq 3,\lambda}(\bm{k})&=&0 ~,
\end{eqnarray}
where $\bm{k}_{2} = \bm{k} - \bm{k}_{1}$ in Eq.~(\ref{eq: TCA0 pi}).
In the tight-coupling limit, the relative velocity between photons and
baryons vanishes as well as in the first-order cosmological perturbation theory.
However, the anisotropic stress of photons is present due to the
quadratic of the photon velocity.

Second, the solutions at first-order in the tight-coupling
expansion are given as 
\begin{widetext}\begin{eqnarray}
\frac{1+R}{R}\delta v^{(2, I)}_{\gamma {\rm b}~\lambda} (\bm{k})
&=& \frac{\sqrt{3}}{20}\left( \frac{k}{\dot{\tau}_{c}}\right)\Delta^{(2)}_{2, \lambda}(\bm{k})
-\left( \frac{\mathcal{H}}{\dot{\tau}_{c}}\right)\left( \omega^{(2)}_{\lambda}(\bm{k})+v^{(2)}_{{\rm b}~\lambda}(\bm{k})\right) \nonumber \\
&&+\int{\frac{d^{3}k_{1}}{(2\pi)^{3}}}\left[ -2\delta v^{(1)}_{\gamma {\rm b}~0}(k_{1})(\delta^{(1)}_{\rm b}+\delta^{(1)}_{\gamma}+\Psi^{(1)})(k_{2}) \right] \sqrt{\frac{4\pi}{3}}Y^{*}_{1 ,\lambda}(\bm{\hat{k}}_{1}) \nonumber \\
&&+\int{\frac{d^{3}k_{1}}{(2\pi)^{3}}}\left[ \frac{1}{2}v^{(1)}_{{\rm b}~0}(k_{1})\Pi^{(1)}_{\gamma~0} (k_{2}) \right] \sqrt{\frac{4\pi}{3}}Y^{*}_{1 ,\lambda}(\bm{\hat{k}}_{1}) \nonumber \\
&&+\int{\frac{d^{3}k_{1}}{(2\pi)^{3}}} \left[ 2v^{(1)}_{\gamma~0}(k_{1})\delta v^{(1)}_{\gamma {\rm b}~0}(k_{2}) \right]\sqrt{\frac{4\pi}{3}}Y^{*}_{1,\lambda}(\bm{\hat{k}}_{1}) \nonumber \\
&&+\frac{1}{R}\int{\frac{d^{3}k_{1}}{(2\pi)^{3}}}\left[ -2\delta v^{(1)}_{\gamma {\rm b}~0}(k_{1})(\delta^{(1)}_{\gamma}+\Psi^{(1)})(k_{2})\right]\sqrt{\frac{4\pi}{3}}Y^{*}_{1, \lambda}(\bm{\hat{k}}_{1}) \nonumber \\
&&+\int\frac{d^{3}k_{1}}{(2\pi)^{3}} \left( \frac{k_{1}}{\dot{\tau}_{c}}\right) \left[ -\frac{1}{2}\delta^{(1)}_{\gamma}(k_{1})(\Psi^{(1)}+\Phi^{(1)})(k_{2}) -2\Psi^{(1)}(k_{1})\delta^{(1)}_{\gamma}(k_{2}) \right] \sqrt{\frac{4\pi}{3}}Y^{*}_{1,\lambda}(\bm{\hat{k}}_{1})\nonumber \\
&&+\int{\frac{d^{3}k_{1}}{(2\pi)^{3}}} \left( \frac{k_{2}}{\dot{\tau}_{c}}\right)\left[ \frac{1}{2} v^{(1)}_{\gamma~0}(k_{1})\left( \delta^{(1)}_{\gamma}+4\Psi^{(1)} \right)(k_{2}) \right]\sqrt{\frac{4\pi}{3}}Y^{*}_{1,\lambda}(\bm{\hat{k}}_{1}) \nonumber \\
&&+\int{\frac{d^{3}k_{1}}{(2\pi)^{3}}}\left( \frac{k_{1}}{\dot{\tau}_{c}}\right)\left[ -\frac{8}{3} v^{(1)}_{\gamma~0}(k_{1})\delta^{(1)}_{\gamma}(k_{2}) \right]\sqrt{\frac{4\pi}{3}}Y^{*}_{1,\lambda}(\bm{\hat{k}}_{1}) \nonumber \\
&&+\int{\frac{d^{3}k_{1}}{(2\pi)^{3}}}\left( \frac{k_{2}}{\dot{\tau}_{c}}\right) \left[ -\frac{2}{3} v^{(1)}_{{\rm b}~0}(k_{1})v^{(1)}_{{\rm b}~0}(k_{2}) \right] \sqrt{\frac{4\pi}{3}}Y^{*}_{1,\lambda}(\bm{\hat{k}}_{1}) \nonumber \\
&&+ \int\frac{d^{3}k_{1}}{(2\pi)^{3}} \left[ -\frac{15}{4}\Pi^{(1)}_{\gamma~0}(k_{1})v^{(1)}_{{\rm b}~0}(k_{2}) \right] \mathcal{Y}^{2,1}_{1,\lambda}(\bm{\hat{k}}_{1},\bm{\hat{k}}_{2})\nonumber \\
&&+\frac{1}{R}\int{\frac{d^{3}k_{1}}{(2\pi)^{3}}}\left[ -\frac{5}{2}\Pi^{(1)}_{\gamma~0}(k_{1})v^{(1)}_{{\rm b}~0}(k_{2})\right] \mathcal{Y}^{2,1}_{1,\lambda}(\bm{\hat{k}}_{1},\bm{\hat{k}}_{2}) \nonumber \\
&&+\int{\frac{d^{3}k_{1}}{(2\pi)^{3}}}\left( \frac{k_{1}}{\dot{\tau}_{c}}\right) \left[ -\frac{10}{3} v^{(1)}_{{\rm b}~0}(k_{1})v^{(1)}_{{\rm b}~0}(k_{2}) \right] \mathcal{Y}^{2,1}_{1,\lambda}(\bm{\hat{k}}_{1},\bm{\hat{k}}_{2}) ~, \label{eq: CPT2TCA1 ell1}
\end{eqnarray}
\begin{eqnarray}
\frac{9}{10}\Delta^{(2,I)}_{2 ,\lambda} (\bm{k})
&=&-\left( \frac{k}{\dot{\tau}_{c}}\right)\frac{\sqrt{3}}{3}\Delta^{(2)}_{1, \lambda}(\bm{k}) \nonumber \\
&&+\int{\frac{d^{3}k_{1}}{(2\pi)^{3}}}\left[ -9\Pi^{(1)}_{\gamma~0}(k_{1}) (\delta^{(1)}_{\rm b}+\Psi^{(1)})(k_{2}) \right] \sqrt{\frac{4\pi}{5}}Y^{*}_{2,\lambda}(\bm{\hat{k}}_{1}) \nonumber \\
&&+\int\frac{d^{3}k_{1}}{(2\pi)^{3}} \left( \frac{k_{1}}{\dot{\tau}_{c}}\right)\left[ -\frac{16}{3} v^{(1)}_{\gamma~0}(k_{1})(\Psi^{(1)}+\Phi^{(1)})(k_{2}) \right] \sqrt{\frac{4\pi}{5}}Y^{*}_{2,\lambda}(\bm{\hat{k}}_{1})\nonumber \\
&&+\int{\frac{d^{3}k_{1}}{(2\pi)^{3}}}\left[ 18v^{(1)}_{\gamma~0}(k_{1})v^{(1)}_{\gamma~0}(k_{2}) +8v^{(1)}_{\gamma~0}(k_{1})\delta v^{(1)}_{\gamma {\rm b}~0}(k_{2}) \right] \mathcal{Y}^{1,1}_{2,\lambda}(\bm{\hat{k}}_{1},\bm{\hat{k}}_{2}) \nonumber \\
&&+ \int{\frac{d^{3}k_{1}}{(2\pi)^{3}}} \left( \frac{k_{1}}{\dot{\tau}_{c}}\right)\left[ \left( 10\delta^{(1)}_{\gamma}(k_{1})-8\Phi^{(1)}\right)(k_{1})v^{(1)}_{\gamma~0}(k_{2}) \right]\mathcal{Y}^{1,1}_{2,\lambda}(\bm{\hat{k}}_{1},\bm{\hat{k}}_{2}) ~, \label{eq: CPT2TCA1 ell2}\\
\Delta^{(2,I)}_{3,\lambda}(\bm{k})&=&-\left( \frac{k}{\dot{\tau}_{c}}\right)\frac{2\sqrt{2}}{5}\Delta^{(2)}_{2,\lambda}+15\int\frac{d^{3}k_{1}}{(2\pi)^{3}} \left[ \Pi^{(1)}_{\gamma~0}(k_{1})v^{(1)}_{\gamma~0}(k_{2}) \right] \mathcal{Y}^{2,1}_{3,\lambda}(\bm{\hat{k}}_{1},\bm{\hat{k}}_{2}) ~, \label{eq: CPT2TCA1 ell3}\\
\Delta^{(2,I)}_{\ell\geq 4,\lambda}(\bm{k}) &=& 0 ~,
\end{eqnarray}\end{widetext}
where $R\equiv 3\rho^{(0)}_{\rm b}/(4\rho^{(0)}_{\gamma})$.
To derive the above solutions, the time derivative of the first-order
curvature perturbation $\dot{\Phi}$ is ignored due to the conservation
of the curvature perturbation on large scales.
In the right-hand side of Eqs.~(\ref{eq: CPT2TCA1 ell1}), (\ref{eq: CPT2TCA1
ell2}), and (\ref{eq: CPT2TCA1 ell3}), for simplicity, we omit the
superscript for the order of the tight-coupling parameter.
Note that the octopole does not vanish and higher multipoles than $\ell
= 3$ are equal to zero at this order \cite{Saga:2014jca}.
\subsection{Einstein equation of the vector mode}
In this subsection, we derive the second-order Einstein equation and hereafter we focus on the vector mode only.
The evolution equation for the metric perturbation of the vector mode,
i.e., $\omega_{\lambda}$ can be derived from 
the $(i,j)$ component of the Einstein equation as 
\begin{eqnarray}
a^{2}G^{i}{}_{j} &=&
e^{2\Phi}\left( \Phi^{,i}{}_{,j}-\Psi^{,i}{}_{,j}\right) + \Phi^{,i}\Phi_{,j}-\Psi^{,i}\Psi_{,j}-\left( \Phi^{,i}\Psi_{,j}+\Phi_{,j}\Psi^{,i}\right) \notag \\
&&+\mathcal{H}\left[ \dot{\chi}^{i}{}_{j}-\left( \omega^{i}{}_{,j}+\omega_{j}{}^{,i}\right)\right] +\frac{1}{2}\left[ \ddot{\chi}^{i}{}_{j}-\left( \dot{\omega}^{i}{}_{,j}+\dot{\omega}_{j}{}^{,i}\right)-\chi^{i}{}_{j}{}^{,a}{}_{,a} \right] \notag \\
&&+ \mbox{(diagonal part)}~\delta^{i}{}_{j}~, \label{Einstein t}
\end{eqnarray}
and 
\begin{eqnarray}
T_{\rm r}^{i}{}_{j}&=&\rho_{\rm r}\Pi_{\rm r}^{i}{}_{j}+\mbox{(diagonal part)}~\delta^{i}{}_{j} ~, \\
T_{\rm m}^{i}{}_{j}&=&\rho_{\rm m}v^{(1)}_{{\rm m} i}v^{(1)}_{{\rm m} j}+\mbox{(diagonal part)}~\delta^{i}{}_{j} ~, \label{em t}
\end{eqnarray}
where $T_{\rm r}^{i}{}_{j}$ and $T_{\rm m}^{i}{}_{j}$ represent the
energy-momentum tensors of massless
(relativistic) particles such as photons and neutrinos, and massive
(nonrelativistic) particles such as baryons and dark matter, respectively.
The equation of the vector mode is derived by acting the
projection operator, $-O^{(-\lambda)}_{ij}(\bm{\hat{k}})$ which can pull the vector mode, as
\begin{widetext}\begin{eqnarray}
\dot{\omega}^{(2)}_{\lambda}(\bm{k})+2\mathcal{H}\omega^{(2)}_{\lambda}(\bm{k})
&=& \frac{2}{5\sqrt{3}}\frac{1}{k}\left( 8\pi Ga^{2}\rho^{(0)}_{\gamma}\Delta^{(2)}_{2,\lambda}(\bm{k})+8\pi Ga^{2}\rho^{(0)}_{\nu}\mathcal{N}^{(2)}_{2,\lambda}(\bm{k})\right) \notag \\
&& +\int\frac{d^{3}k_{1}}{(2\pi)^{3}}4k_{1}\left[ \Phi^{(1)}(k_{1})\Psi^{(1)}(k_{2}) \right] \sqrt{\frac{4\pi}{3}}Y^{*}_{1,\lambda}(\bm{\hat{k}}_{1}) \notag \\
&& -\int\frac{d^{3}k_{1}}{(2\pi)^{3}}\frac{4}{\sqrt{3}}\frac{k^{2}_{1}}{k}\left[ \Phi^{(1)}(k_{1})\Phi^{(1)}(k_{2})+\Psi^{(1)}(k_{1})\Psi^{(1)}(k_{2}) \right] \sqrt{\frac{4\pi}{5}}Y^{*}_{2,\lambda}(\bm{\hat{k}}_{1}) \notag \\
&& +\sum_{{\rm s} = {\rm b}, {\rm dm}}8\pi Ga^{2} \rho^{(0)}_{\rm s}\int{\frac{d^{3}k_{1}}{(2\pi)^{3}}} \left[ \frac{4}{k}v^{(1)}_{{\rm s}~0}(k_{1})v^{(1)}_{{\rm s}~0}(k_{2}) \right]\sqrt{\frac{4\pi}{3}}Y^{*}_{1,0}(\bm{\hat{k}}_{1})\sqrt{\frac{4\pi}{3}}Y^{*}_{1,\lambda}(\bm{\hat{k}}_{2}) ~. \label{eq: metric}
\end{eqnarray}\end{widetext}
Note that in the Poisson gauge, the vector metric perturbation is only included in $\omega_{i}$.
When we consider the evolution equation up to the first order in the standard cosmology with perfect fluids, the right-hand side of Eq.~(\ref{eq: metric}) becomes zero.
As a result, the vector mode has only a decaying
solution, which is neglected in the linear theory.
\section{Cosmological Magnetic Fields}
In this section, we show the evolutions and spectra of magnetic fields
driven by the Harrison mechanism. 
In the previous studies
\cite{Ichiki:2007hu,2011MNRAS.414.2354F,Nalson:2013jya,Matarrese:2004kq,Siegel:2006px},
generated magnetic fields are partially estimated by numerical or
analytical ways, and there is a small discrepancy
between in Refs.~\cite{Ichiki:2007hu} and \cite{2011MNRAS.414.2354F}.
In this paper, we build on Ref.~\cite{Ichiki:2007hu} and
expand the work by including all contributions numerically. 
The source terms of the magnetic fields consist of three
contributions, i.e., $\delta^{(1)}_{\gamma}\delta v^{(1)}_{\gamma {\rm
b}}$, $v^{(1)}_{\rm b}\Pi^{(1)}_{\gamma}$, and $\delta v^{(2)}_{\gamma
{\rm b}}$, which hereafter we call ``the slip term'', ``the anisotropic
stress term'' and ``the second-order slip term'', respectively.
 
In this paper, we focus on three issues on the generation of
magnetic fields at recombination. First,
we consider how large is the contribution of the second-order slip term
on magnetic fields compared with the contributions of the slip and the
anisotropic stress terms. Second, to evaluate the total spectrum
of magnetic fields we
need to include the cross-correlation terms between the sources,
namely, $P_{B}\sim \left<\left( B_{\rm 2nd~Slip} + B_{\rm Slip} + B_{\rm
Anis}\right)^{2}\right>$. The cross terms can be negative and it has the
possibility to cancel the generated magnetic fields from each
of the source terms. Third, we try to find the cause of the small
discrepancy between 
Refs.~\cite{Ichiki:2007hu} and \cite{2011MNRAS.414.2354F}.
In the following subsections, we show the evolutions and spectra of
magnetic fields and the answers of the above three considerable questions.

\subsection{Configuration in Fourier space}\label{sec: anti-sym vec}
Before moving to the results, we mention the
configuration of Fourier space in carrying out the
convolution since all second-order quantities are written by the products of the first-order scalar perturbations.
To calculate the second-order power spectrum, we decompose the second-order variable into the transfer function and the primordial amplitude \cite{Pettinari:2014vja} as
\begin{equation}
\Delta^{(2)}(\eta, \bm{k}) = \int{\frac{d^{3}k_{1}}{(2\pi)^{3}}}\int{\frac{d^{3}k_{2}}{(2\pi)^{3}}}\delta(\bm{k}-\bm{k}_{1}-\bm{k}_{2}) \Delta^{(2)}_{\rm T}(\eta, \bm{k},\bm{k}_{1},\bm{k}_{2})\Phi(\bm{k}_{1})\Phi(\bm{k}_{2}) ~, \label{eq: gen transfer}
\end{equation}
where $\Delta^{(2)}_{\rm T}$ and $\Phi$ are the second-order transfer function and the primordial amplitude, respectively.
The ensemble average of the variance of the primordial amplitude
can be expressed as $\Braket{\Phi^{*}(\bm{k}_{1})\Phi(\bm{k}_{2})} =
(2\pi)^{3}P_{\Phi}(k_{1})\delta(\bm{k}_{1}-\bm{k}_{2})$, where
$P_{\Phi}(k)$ is the primordial power spectrum determined from cosmological
observations such as CMB and large scales structure. 
In this paper, we use the power-law spectrum as 
\begin{equation}
\frac{k^{3}}{2\pi^{2}}P_{\Phi}(k) = \frac{4}{9}\mathcal{A}_{s}\left( \frac{k}{k_{0}}\right)^{n_{s}-1} ~,
\end{equation}
where the parameters, $\mathcal{A}_{s}$ and $n_{s}$, are the amplitude and the spectral index of primordial perturbations, respectively.
We set $\mathcal{A}_{s} = 2.4\times 10^{-9}$ from the WMAP nine-year results \cite{2013ApJS..208...20B}, and for simplicity, we consider a scale-invariant spectrum, namely, $n_{s} = 1.0$.

In this decomposition, the second-order power spectrum can be written as
\begin{equation}
P_{\Delta^{(2)}}(\eta ,\bm{k}) = 2\int{\frac{d^{3}k_{1}}{(2\pi)^{3}}}\left[ \Delta^{(2)}_{\rm T}(\eta,\bm{k},\bm{k}_{1},\bm{k}_{2})\right]^{2} P_{\Phi}(k_{1}) P_{\Phi}(k_{2}) ~, \label{eq: gen power}
\end{equation}
where we can use the symmetry under the exchange of $k_{1}$ and $k_{2}$ without loss of generality to derive the above relation.
Note that $\bm{k}_{2} = \bm{k}-\bm{k}_{1}$ should be satisfied implicitly, namely, $\Delta^{(2)}_{\rm T}(\eta, \bm{k}, \bm{k}_{1}, \bm{k}_{2}) = \Delta^{(2)}_{\rm T}(\eta, \bm{k}, \bm{k}_{1}, \bm{k}-\bm{k}_{1})$ and $P_{\Phi}(k_{2}) = P_{\Phi}(\left| \bm{k}-\bm{k}_{1}\right|)$, as is mentioned before.

We need to solve the Einstein-Boltzmann system in
$(k, k_{1},k_{2})$ space.
Note that the transfer function is transformed under the rotation of
$\phi$ as
\begin{equation}
\Delta^{(2)}_{\rm T}(\eta, \bm{k},\bm{k}_{1},\bm{k}_{2}) \longrightarrow \Delta^{(2)}_{\rm T} (\eta, \bm{k},\bm{k}_{1},\bm{k}_{2})e^{im\phi} ~.
\end{equation}
In practice, we take $\phi = \theta = 0$, $\phi_{1} = 0$, and $\phi_{2} = \pi$ for $\bm{k}$, $\bm{k}_{1}$, and $\bm{k}_{2}$, respectively.
In other words, the transfer function under the exchange of $k_{1}$ and $k_{2}$ transforms as
\begin{equation}
\Delta^{(2)}_{\rm T}(\eta, \bm{k},\bm{k}_{1},\bm{k}_{2}) = (-1)^{m}\Delta^{(2)}_{\rm T}(\eta, \bm{k},\bm{k}_{2},\bm{k}_{1})~.
\end{equation}
It is very interesting that for the case of the scalar and tensor modes, the dominant contribution comes from near the $k\sim k_{1}\sim k_{2}$.
Because the transfer functions of the scalar ($m=0$) and tensor ($m=\pm 2$) modes are symmetric under the exchange of $k_{1}$ and $k_{2}$.
On the other hand, the vector mode does not have dominant contribution near the $k\sim k_{1}\sim k_{2}$ since the transfer function of the vector mode is antisymmetric under the exchange of $k_{1}$ and $k_{2}$.

In the vector mode, we can naively consider three configurations of
the triangle $\bm{k} = \bm{k}_{1} + \bm{k}_{2}$ that contribute to
the power spectrum on superhorizon scales,
namely, the one where both $k_{1}$ and $k_{2}$ are on
superhorizon scales, where $k_{1}$ ($k_{2}$) is at superhorizon scales
while $k_{2}$ ($k_{1}$) is at subhorizon scales, and where both $k_{1}$
and $k_{2}$ are at subhorizon scales. When both $k_{1}$ and
$k_{2}$ are at subhorizon scales, the square of the primordial
power spectrum in Eq.~(\ref{eq: gen power}) does not contribute on the
power spectrum because of the antisymmetric nature of the vector
transfer function.
In the vector mode, therefore, we need a careful treatment of how to
sample wave numbers in $(k,k_{1},k_{2})$ space.
This difficulty does not arise in the second-order scalar and
tensor modes calculations because the transfer functions of the
scalar and tensor modes have a symmetry under the exchange of $k_{1}$ and $k_{2}$.
\subsection{Evolutions of magnetic fields}\label{sec: evolve mag}
\begin{figure}[t]
\begin{center}
\rotatebox{0}{\includegraphics[width=0.45\textwidth]{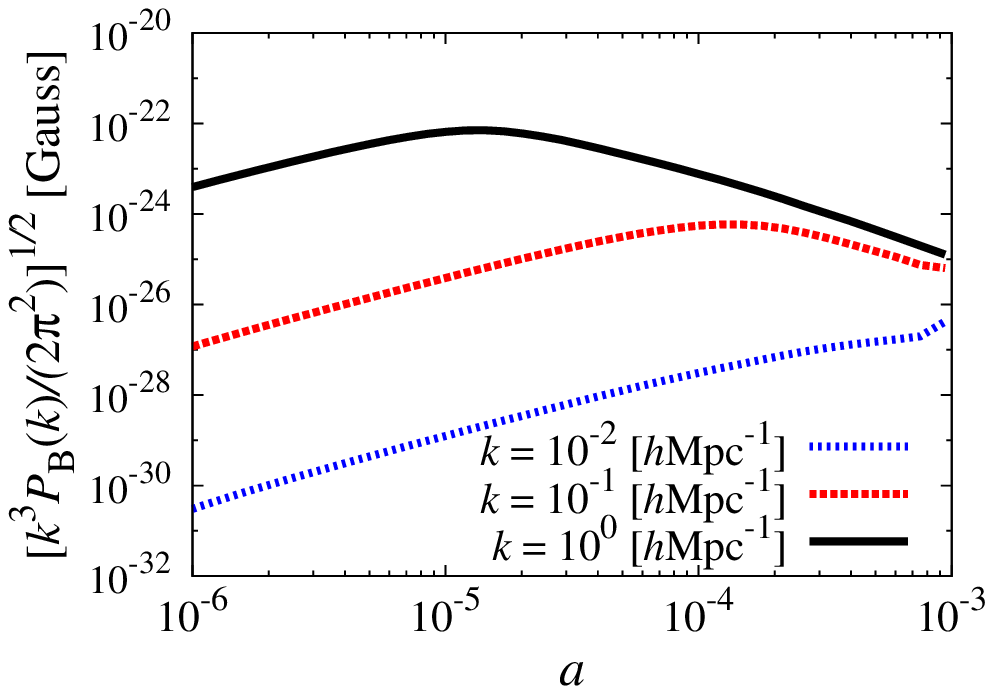}}
\rotatebox{0}{\includegraphics[width=0.45\textwidth]{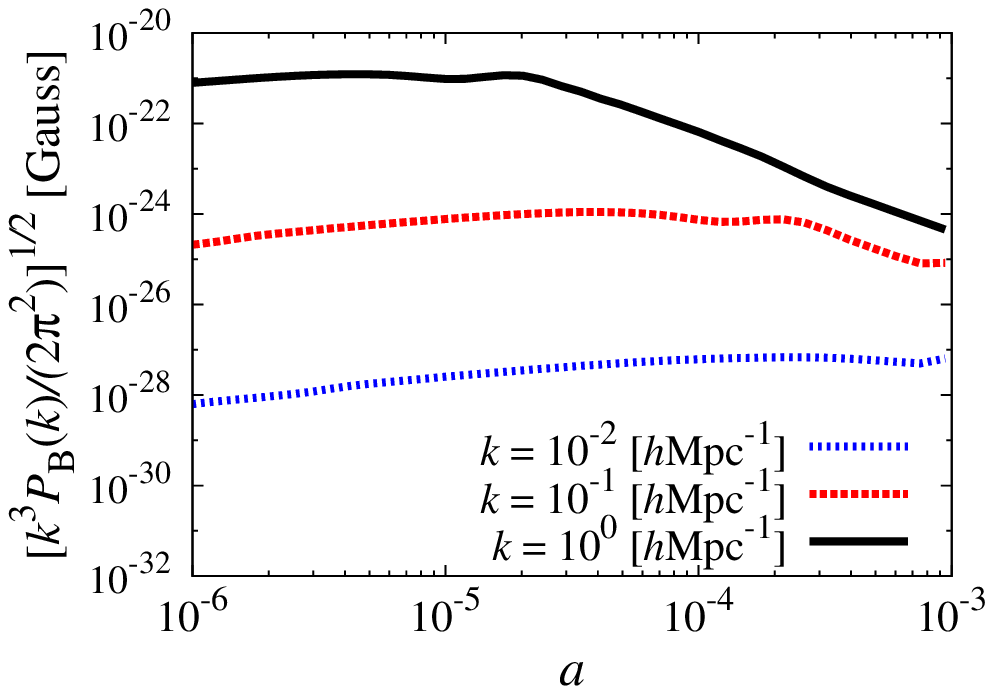}}
\rotatebox{0}{\includegraphics[width=0.45\textwidth]{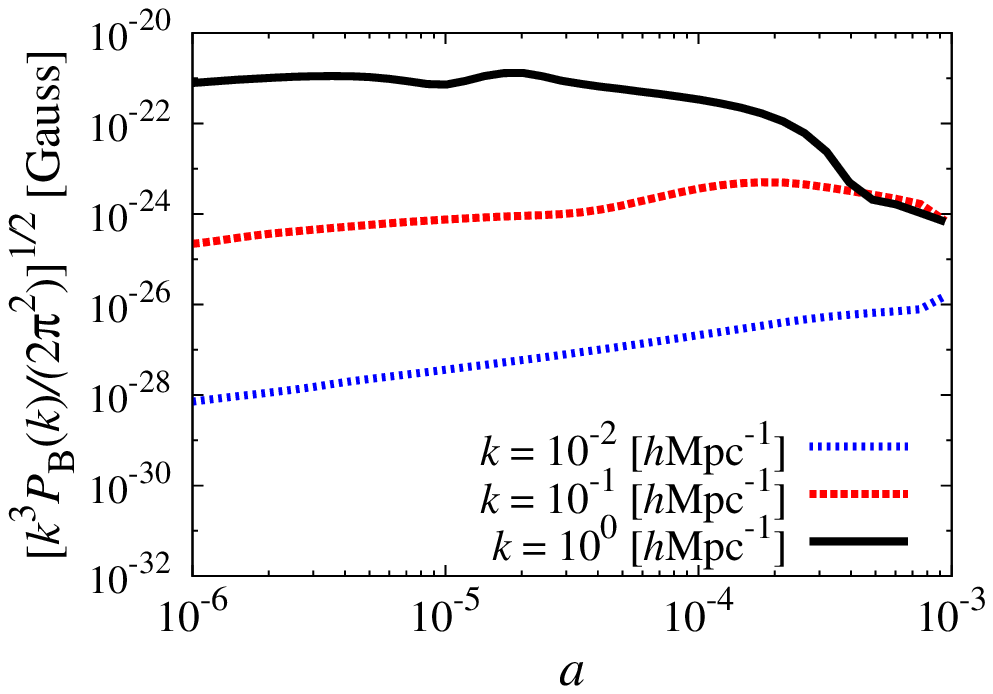}}
\end{center}
\caption{%
Evolutions of generated magnetic fields sourced by the slip term ({\it top left}), the anisotropic stress term ({\it top right}), and the second-order slip term ({\it bottom}).
We show evolutions of generated magnetic fields for wavenumbers $k = 10^{-2}~h{\rm Mpc}^{-1}$, $10^{-1}~h{\rm Mpc}^{-1}$, and $10^{0}~h{\rm Mpc}^{-1}$ as indicated in the above panels.
We can see that magnetic fields at smaller scales generated earlier and their amplitudes are larger.
}
\label{fig:Evolution}
\end{figure}

In Fig.~\ref{fig:Evolution}, we show the evolutions of magnetic fields induced by the slip, the anisotropic stress, and the second-order slip terms.
First, we focus on the evolutions of magnetic fields before the horizon crossing.
The time evolutions of magnetic fields can be approximated by a
power law, and 
the powers of the slip and the anisotropic stress terms are $B_{\rm
Slip}\propto \eta^{1.5}$ and $B_{\rm Anis}\propto \eta^{0.5}$, respectively.
These results correspond to the previous study \cite{Ichiki:2007hu}.
Furthermore, the second-order slip term is proportional to $\eta^{0.5}$
on superhorizon scales. 
We can explain the coincidence of the powers between the anisotropic
stress and the second-order slip terms as follows.
In Eq.~(\ref{eq: CPT2TCA1 ell1}), the dominant terms in early
times can be
 estimated by using the superhorizon solutions at linear order.
At first glance, the term $\delta^{(1)}_{\gamma}\times \Phi^{(1)}$ in
Eq.~(\ref{eq: CPT2TCA1 ell1}) seems to give a dominant
contribution to the second-order slip term. 
However, because both $\delta^{(1)}_{\gamma}$ and $\Phi^{(1)}$ are
scale invariant in the Poisson gauge and according to the formula of spherical harmonics,
$k_{1}\sqrt{\frac{4\pi}{3}}Y^{*}_{1,m}(\bm{\hat{k}}_{1})+k_{2}\sqrt{\frac{4\pi}{3}}Y^{*}_{1,m}(\bm{\hat{k}}_{2}) = k\delta_{m,0}$,
this scale invariant term vanishes in the vector mode.
As a result, we find that the most dominant term in Eq.~(\ref{eq: CPT2TCA1 ell1}) is that proportional to $v^{(1)}_{\rm
b}\Pi^{(1)}_{\gamma}$, which is the same form as the anisotropic stress term.
Therefore, the powers of the time evolutions of the anisotropic
stress and second-order slip terms coincide. 

Next, we discuss the evolutions of magnetic fields after the horizon
crossing. We can see that the magnetic fields from the slip and the
anisotropic stress terms start to decay adiabatically as $\propto
a^{-2}$, after the horizon crossing since their sources also diminish
after the horizon crossing.  On the other hand, magnetic fields induced
by the second-order slip term do not
decay adiabatically even after the horizon crossing.
This arises from the fact that the second-order
slip term $\delta v^{(2)}_{\gamma {\rm b}}$ continues to grow even after the source terms
from the first-order perturbations become negligible, until the corresponding scale reaches
the Silk damping scale. Therefore, the purely second-order
perturbations can contribute to the magnetic
field generation even if the product of the first-order
perturbations is absent.  When we neglect the product of the
first-order perturbations, the evolution equations for magnetic fields
are corresponding to the case of the first-order magnetic fields
generation \cite{2012PhRvD..85d3009I}.  The relative velocity between
photons and baryons, $\delta v^{(2)}_{\gamma{\rm b}~\lambda}$
contributes to the generation of magnetic fields after the horizon
crossing.  This additional enhancement can be seen in the bottom of
Fig.~\ref{fig:Evolution} at $k=10^{0}~h{\rm Mpc}^{-1}$.  However,
magnetic fields induced by the additional enhancement undergo nontrivial cancellation
after the Silk damping epoch and magnetic fields decay faster than
the adiabatic decay that it is proportional to $a^{-2}$ \cite{2012PhRvD..85d3009I}.  The final amplitude of magnetic
fields is consequently determined by the initial amplitude around
horizon crossing.

\subsection{Spectra of magnetic fields}
\begin{figure}[t]
\begin{center}
\rotatebox{0}{\includegraphics[width=0.45\textwidth]{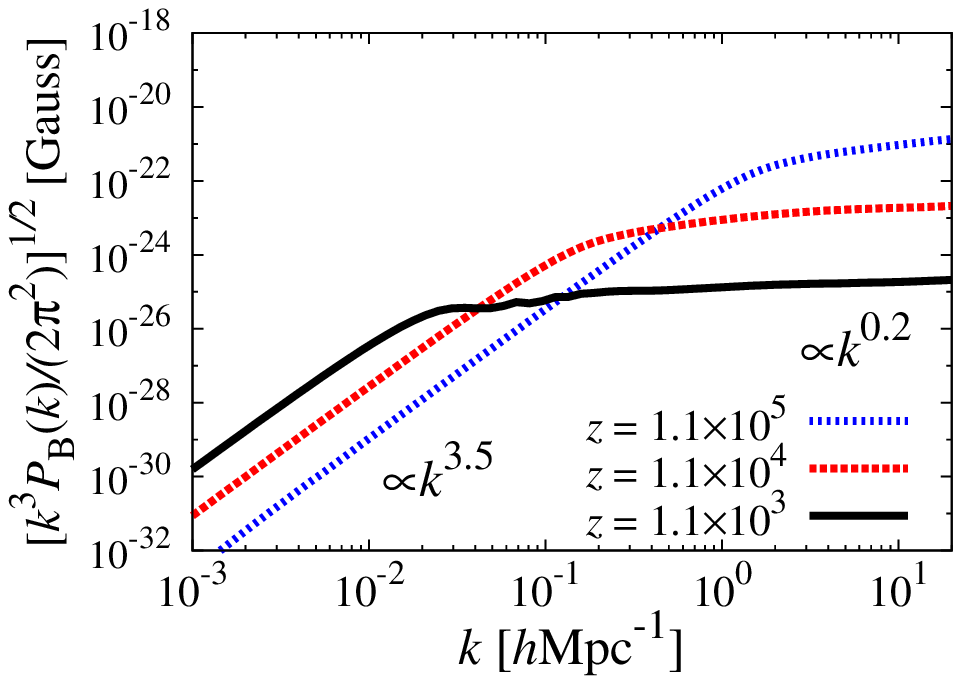}}
\rotatebox{0}{\includegraphics[width=0.45\textwidth]{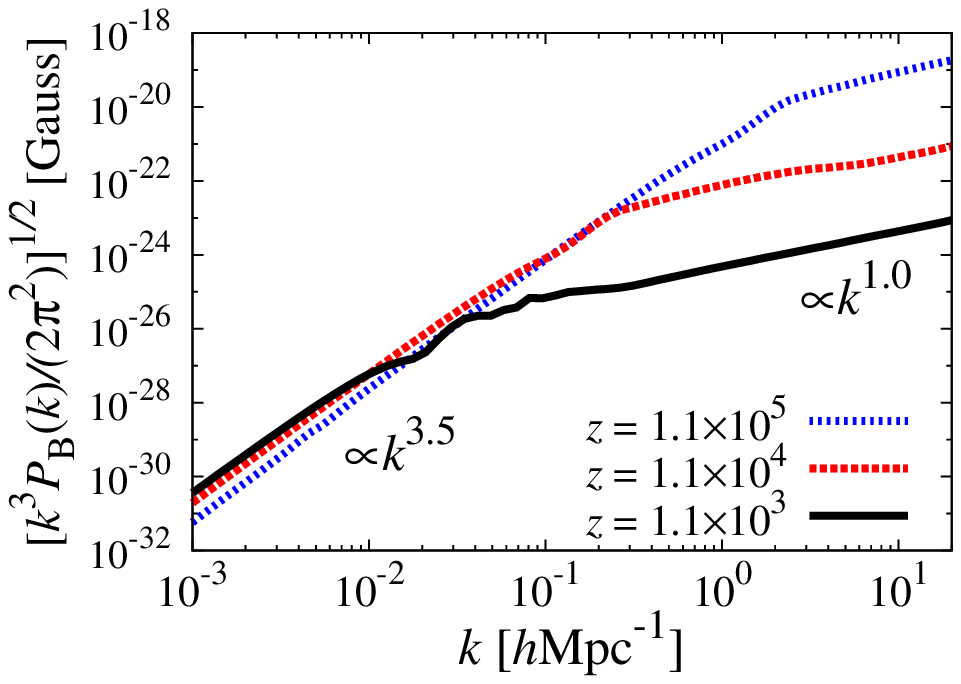}}
\rotatebox{0}{\includegraphics[width=0.45\textwidth]{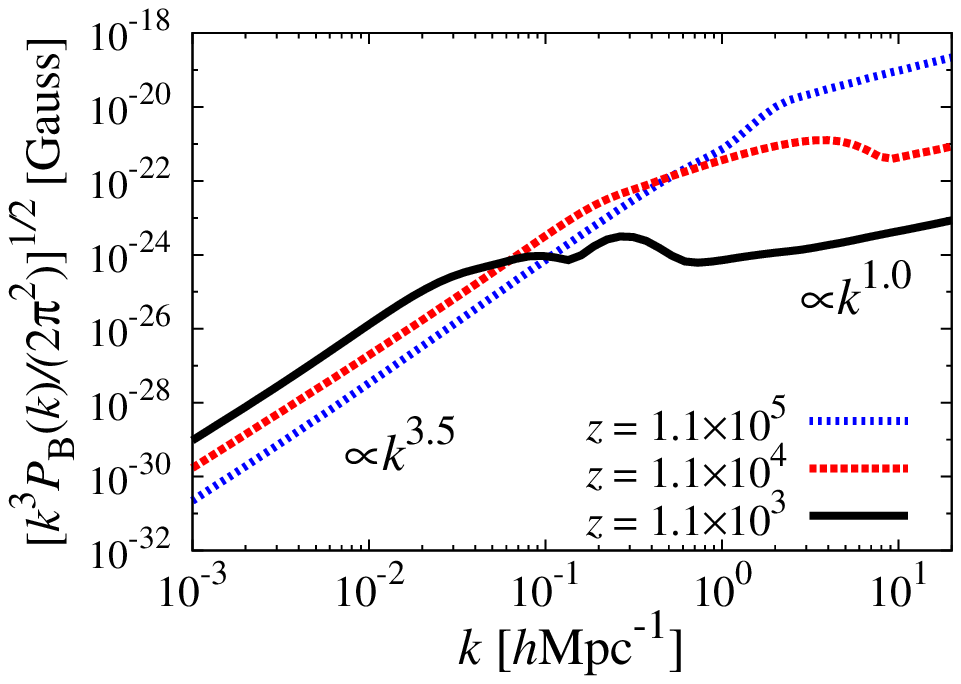}}
\end{center}
\caption{%
Spectra of generated magnetic fields sourced by the slip term ({\it top left}), the anisotropic stress term ({\it top right}), and the second-order slip term ({\it bottom}).
We show spectra of generated magnetic fields for redshifts $z = 1.1\times 10^{5}$, $1.1\times 10^{4}$, and $1.1\times 10^{3}$ as indicated in the above panels.
}
\label{fig: Spectra}
\end{figure}

Next, we show the spectra of magnetic fields induced by
the slip, the anisotropic stress, and the second-order slip terms in
Fig.~\ref{fig: Spectra}. 
From Fig.~\ref{fig: Spectra},
resultant magnetic fields are dominated by the anisotropic stress and second-order slip terms on smaller scales, i.e., $k\gtrsim 1.0~h{\rm Mpc}^{-1}$, at $1+z = 1100$.
Conversely, on these scales, the slip term is a subdominant source for magnetic fields.

On superhorizon scales, we can
see that the spectra of magnetic fields are proportional to $k^{3.5}$, which also corresponds to the results about the slip and the anisotropic stress terms in Ref.~\cite{Ichiki:2007hu}.
This power is also consistent with the power spectrum for causal magnetic fields \cite{Durrer:2003ja}.
From Ref.~\cite{Ichiki:2007hu}, the magnetic power on superhorizon scales can be estimated as below.
For example, we focus on magnetic fields induced by the slip term.
We can integrate the evolution equation for the second-order magnetic fields (\ref{eq: vec mag}) and take the ensemble average as
\begin{equation}
\left. \frac{k^{3}}{2\pi^{2}} P_{B}(k)\right|_{\rm Slip} \propto k^{5} \int{\frac{d^{3}k_{1}}{(2\pi)^{3}}}
\left( 1-\mu^{2}_{1}\right)
P_{\Phi}(k_{1})P_{\Phi}(k_{2})
\left[ S^{2}(k_{2},k_{1})-\frac{k_{1}}{k_{2}}S(k_{1},k_{2})S(k_{2},k_{1})\right] ~, \label{eq: power on sh}
\end{equation}
where $\mu_{1} = \cos\theta_{1}$ and $S(k_{1},k_{2})$ is defined as
\begin{equation}
S(k_{1},k_{2}) = \int{d\eta}a^{2}(\eta)\rho^{(0)}_{\gamma}(\eta)\delta^{(1)}_{\gamma}(k_{1}, \eta)\delta v^{(1)}_{\gamma {\rm b}~0}(k_{2},\eta) ~,
\end{equation}
where for simplicity, we omit the time dependence of $S(k_{1},k_{2})$.
To proceed the estimation of the magnetic power, we take the limit $k/k_{1}\to 0$.
This approximation can include the contributions from subhorizon scales.
In this limit,
\begin{equation}
k_{2} = k_{1}\left[ 1 -\frac{k}{k_{1}}\mu_{1} + O\left( (k/k_{1})^{2} \right) \right] ~.
\end{equation}
Furthermore, we can approximate the integrated source term as $S(k_{1},k_{2})\approx S(k_{2},k_{1})\approx T(k_{1})$ since $S(k_{1},k_{2})$ can be treated as $k$ independent in the above limit.
Then by using the fact that $P_{\Phi}(k)\propto k^{n_{s}-4}$, Eq.~(\ref{eq: power on sh}) can be rewritten as
\begin{eqnarray}
\left. \frac{k^{3}}{2\pi^{2}} P_{B}(k)\right|_{\rm Slip} 
&\propto& k^{5} \int k^{2}_{1} dk_{1} \int d\mu_{1} \left( 1-\mu^{2}_{1}\right) k^{n_{s}-4}_{1}k^{n_{s}-4}_{2} \left( 1-\frac{k_{1}}{k_{2}}\right) T^{2}(k_{1}) \\
&\propto& k^{7}~.
\end{eqnarray}
This nonlinear power law can be seen in Fig.~\ref{fig: Spectra}.
Note that if we use the superhorizon solution only, namely, $\delta^{(1)}_{\gamma}\propto (k\eta)^{0}$ and $\delta v^{(1)}_{\gamma {\rm b}~0}\propto k^{3}\eta^{5}$, the magnetic power is returned as $\propto k^{8}$.
This superhorizon power does not match for our numerical results.

By using our numerical code,
we trace a possible cause of the discrepancies
between the results in previous studies. 
As we noted in Sec.~\ref{sec: anti-sym vec}, the transfer function
of the vector mode is antisymmetric under the exchange of $k_{1}$ and
$k_{2}$, and therefore, the isosceles configuration such that $k_{1} =
k_{2}$ in Fourier space does not contribute in the calculation of the
power spectrum. 
However, to achieve the result correctly, 
contributions from the configurations of 
$k_{1}\lesssim k_{2}$ and $k_{2}\lesssim k_{1}$ should be included. 
When these contributions are not included in the numerical calculation, the power spectrum of magnetic
fields on superhorizon scales shows $\propto k^{4}$, which corresponds to the result
obtained in Ref.~\cite{2011MNRAS.414.2354F}.

On subhorizon scales, we can find that the spectra of magnetic fields induced by the slip,
the anisotropic stress, and the 
second-order slip terms are proportional to $k^{0.2}$, $k^{1.0}$, and $k^{1.0}$,
respectively.
The spectra induced by the slip and the anisotropic stress terms are consistent with Ref.~\cite{Ichiki:2007hu}.
A noticeable feature in the spectrum
induced by the second-order slip term is the additional amplification at
$k\approx 5.0 \times 10^{-1}~h{\rm Mpc}^{-1}$ at $z=1100$.
As discussed in Sec.~\ref{sec: evolve mag}, this additional amplification is due to the second-order relative velocity between photons and baryons after the horizon crossing.
However, this amplification is a temporary effect and the amplified magnetic fields by this effect 
had been erased by the epoch of Silk damping \cite{2012PhRvD..85d3009I}.
Therefore, the amplification cannot be seen for
scales $k\gtrsim 1.0 ~h{\rm Mpc}^{-1}$.

The above discussions and results in this subsection are
valid only for the auto-power spectra of
the magnetic fields from
the slip, the anisotropic stress, and the second-order slip terms. 
In the following subsection, we
focus on the total power spectrum induced by all the
contributions including the cross spectra.
\subsection{The second-order magnetic fields}
\begin{figure}[t]
\begin{center}
\rotatebox{0}{\includegraphics[width=0.65\textwidth]{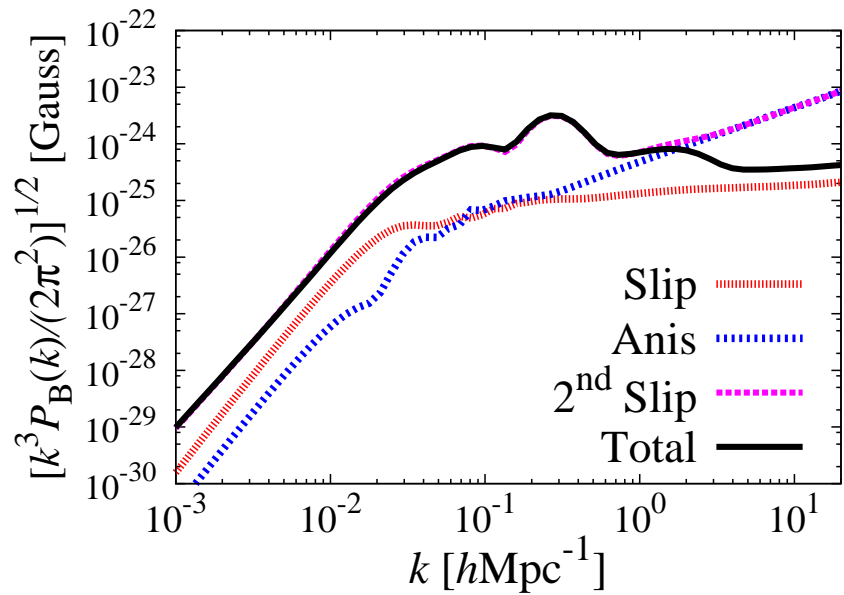}}
\end{center}
\caption{%
Magnetic spectra generated from the slip term, the anisotropic stress term, the second-order slip term, and all terms included the cross terms at recombination ($1+z \simeq 1100$).}
\label{fig: total}
\end{figure}

We depict the total power spectrum at recombination in Fig.~\ref{fig: total}.
It is clear that the second-order slip term gives
a dominant contribution to the total magnetic fields.
The amplitude of magnetic
fields from the second-order slip term is
$10$ times larger than without the second-order slip term. 
However, we find that the magnetic fields from the second-order
slip term are canceled out by the magnetic fields from the anisotropic
stress term on small scales.

By using the tight coupling solution given by Eq.~(\ref{eq: CPT2TCA1 ell1}), this cancellation is easily understood analytically.
As we mentioned in Sec.~\ref{sec: evolve mag},
the dominant term in Eq.~(\ref{eq: CPT2TCA1 ell1}) is the anisotropic stress term
given by
\begin{equation}
\delta v^{(2)}_{\gamma {\rm b}~\lambda} \simeq \frac{1}{1+R}\int{\frac{d^{3}k_{1}}{(2\pi)^{3}}}\left[ -\frac{5}{2}\Pi^{(1)}_{\gamma~0}(k_{1})v^{(1)}_{{\rm b}~0}(k_{2})\right]\mathcal{Y}^{2,1}_{1,\lambda}(\bm{\hat{k}}_{1},\bm{\hat{k}}_{2})~,
\end{equation}
while any other terms related to the anisotropic stress are subdominant with
the baryon-photon ratio $R\propto 3\rho^{(0)}_{\rm
b}/(4\rho^{(0)}_{\gamma})$ as a suppression factor. 
By substituting this expression into Eq.~(\ref{eq: vec mag}), 
the evolution of magnetic fields from the second-order slip
and the anisotropic stress terms is given as
\begin{equation}
\frac{d\mathcal{B}_{\lambda}}{dt} \propto \left[ \frac{5}{4}\frac{1}{1+R} -\frac{5}{4}\right] v^{(1)}_{{\rm b}~0}(k_{1})\Pi^{(1)}_{\gamma~0}(k_{2})\mathcal{Y}^{1,2}_{1,\lambda}(\bm{\hat{k}}_{1},\bm{\hat{k}}_{2}) ~, \label{eq: approx B1}
\end{equation}
where the first and the last terms in the parentheses are coming from the second-order slip term and the anisotropic stress term, respectively.
We can see that the two terms in the parentheses in Eq.~(\ref{eq:
approx B1}) are canceled in the radiation dominated era, where $R$
is negligibly small.
However, in the matter dominated era, the baryon-photon ratio has large
value and this cancellation does not occur. 

The dominant contribution from the second-order slip term is
canceled out by the contribution from the
anisotropic stress term in the radiation
dominated era.
Conversely, there still remain some contributions from the second-order
slip term as shown in
Fig.~\ref{fig: total} and discussed below using the tight-coupling solution.
The subleading contribution from the second-order
slip term can be written as
\begin{equation}
\delta v^{(2)}_{\gamma {\rm b}~\lambda} \simeq \frac{1}{1+R}\int{\frac{d^{3}k_{1}}{(2\pi)^{3}}}\left[ -2\delta v^{(1)}_{\gamma{\rm b}~0}(k_{1})\delta^{(1)}_{\gamma}(k_{2})\right]\sqrt{\frac{4\pi}{3}}Y^{*}_{1,\lambda}(\bm{\hat{k}}_{1})~.
\end{equation}
Then, the evolution equation of magnetic fields induced by the slip and
the second-order slip terms can be rewritten as
\begin{equation}
\frac{d\mathcal{B}_{\lambda}}{dt} \propto \left[ \frac{1}{1+R} +1\right] \delta v^{(1)}_{\gamma{\rm b}~0}(k_{1})\delta^{(1)}_{\gamma}(k_{2}) \sqrt{\frac{4\pi}{3}}Y^{*}_{1,\lambda}(\bm{\hat{k}}_{1})~, \label{eq: approx B2}
\end{equation}
where the first and the last terms in the parentheses are coming from the second-order slip term and the slip term, respectively.
From Eq.~(\ref{eq: approx B2}), we find that the total
amplitude of the spectrum of magnetic fields is twice as large as the case
only with the slip term. This tendency can be seen in Fig.~\ref{fig: total}.

\begin{figure}[t]
\begin{center}
\rotatebox{0}{\includegraphics[width=0.65\textwidth]{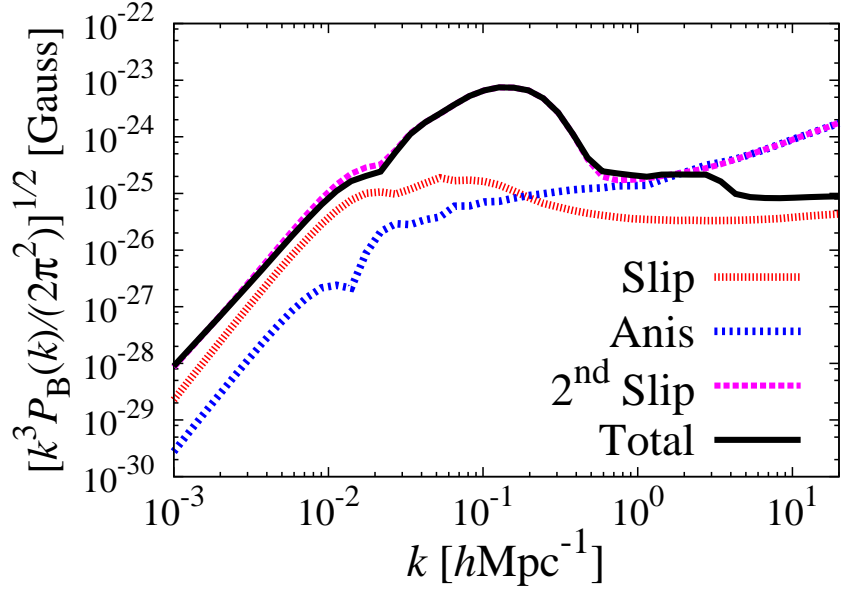}}
\end{center}
\caption{%
Same as Fig.~\ref{fig: total} but for after recombination as ($1+z \simeq 500$).}
\label{fig: total z=500}
\end{figure}

Next, let us discuss the evolution of magnetic fields through the epoch of recombination.
As the process of recombination proceeds, electrons form neutral hydrogen atoms with protons and the number of free electrons rapidly decreases. Accordingly the effect of the Compton scattering on generation of magnetic fields becomes negligible and magnetic fields are no longer generated through the Harrison mechanism.
We show the power spectrum of the second-order magnetic fields after recombination at $z \simeq 500$ in Fig.~\ref{fig: total z=500}.
In Fig.~\ref{fig: total z=500}, one can find new features in the spectrum different from one at recombination on intermediate scales such as $10^{-2}~h{\rm Mpc}^{-1} \lesssim k \lesssim 1.0~h{\rm Mpc}^{-1}$.
These features are nearly consistent with Ref.~\cite{2011MNRAS.414.2354F}.
For instance, magnetic fields induced by the slip term decrease on intermediate scales.
On the other hand, magnetic fields induced by the second-order slip term are enhanced at cosmological recombination since the relative velocity is also enhanced at that era.
We furthermore extend the magnetic spectrum to much smaller scales.
On scales where $1.0~h{\rm Mpc}^{-1} \gtrsim k$, magnetic fields induced by the slip, the anisotropic stress, and the second-order slip terms have the same structures as the magnetic spectrum at recombination shown in Fig.~\ref{fig: total}.
As a result, the spectrum of second-order magnetic fields has a slightly blue tilt on small scales.

\subsection{Nonhelical magnetic fields}
Before closing this section, we investigate the possibility whether
helical magnetic fields are generated in the Harrison mechanism or not
since the helicity may play important roles in
the cosmological observations
\cite{Pogosian:2001np,Caprini:2003vc,Kahniashvili:2005xe,Shiraishi:2013kxa,Shiraishi:2012sn}.
In fact, in Ref.~\cite{Tashiro:2013ita}, the authors
found the evidence of existence of helical magnetic fields on a few Mpc scales.
It is believed that helical magnetic fields can only be generated through the process of parity violation.
Because helicity is conserved in the standard magnetohydrodynamics,
it is a good indicator to probe the generation mechanism of magnetic fields. 
We will show below that the Harrison
mechanism does not induce helical magnetic fields since this mechanism
relies on the standard Compton scattering which does not break the
parity symmetry.

At first, under the existence of helical magnetic fields, the correlation of magnetic fields can be written as
\begin{equation}
\Braket{B_{i}(\bm{k})B_{j}(\bm{k'})} = (2\pi)^{3}\delta(\bm{k}-\bm{k'}) \left[  (\delta_{ij} - \hat{k}_{i}\hat{k}_{j})\frac{P_{B}(k)}{2} + i\epsilon_{ijk}\hat{k}^{k}\frac{P_{H}(k)}{2}\right] ~,
\end{equation}
where $P_{B}(k)$ and $P_{H}(k)$ are the spectra of nonhelical and helical magnetic fields, respectively.
The power spectrum of helical magnetic fields can be pulled by the subtraction as $\Braket{B_{i}(\bm{k})B_{j}(\bm{k'})} -\Braket{B_{j}(\bm{k})B_{i}(\bm{k'})}$.
In the Harrison mechanism that is given by Eq.~(\ref{eq: B_dot}), generated magnetic fields can be symbolically expressed as
\begin{equation}
B_{i}(\bm{k}) \propto \epsilon_{iab}k^{a}\int\frac{d^{3}k_{1}}{(2\pi)^{3}} \hat{k}^{b}_{1} f(\bm{k}_{1},\bm{k}_{2}) X^{(1)}(\bm{k}_{1})Y^{(1)}(\bm{k}_{2}) ~,
\end{equation}
where $f(\bm{k}_{1},\bm{k}_{2})$ is an arbitrary real function of $\bm{k}_{1}$ and $\bm{k}_{2}$, and  $X^{(1)}(\bm{k}_{1})$ and $Y^{(1)}(\bm{k}_{2})$ are the time integrals of first-order scalar perturbations.
The scalar perturbations can be decomposed into the primordial perturbation $\Phi^{(1)}(\bm{k})$ and the transfer function $X_{\rm T}(k_{1})$ or $Y_{\rm T}(k_{2})$ as $X^{(1)}(\bm{k}_{1})Y^{(1)}(\bm{k}_{2}) = \Phi^{(1)}(\bm{k}_{1})\Phi^{(1)}(\bm{k}_{2})X_{\rm T}(k_{1})Y_{\rm T}(k_{2})$.
Note that we use the fact that the purely second-order variables, e.g., $\delta v_{\gamma b i}$, are composed the product of the first-order scalar perturbations.

Finally, we evaluate the helical part of the power spectrum as
\begin{eqnarray}
\Braket{B_{i}(\bm{k})B_{j}(\bm{k'})} -\Braket{B_{j}(\bm{k})B_{i}(\bm{k'})}
& \propto & (2\pi)^{3}\delta(\bm{k}-\bm{k'})
\left( \epsilon_{iab}\epsilon_{ja'b'}-\epsilon_{jab}\epsilon_{ia'b'} \right) k^{a}k^{a'} \notag \\
&& \times \int\frac{d^{3}k_{1}}{(2\pi)^{3}} f(\bm{k}_{1},\bm{k}_{2}) X_{\rm T}(k_{1}) Y_{\rm T}(k_{2})P_{\Phi}(k_{1})P_{\Phi}(k_{2})\notag \\
&& \times \Bigl[  \hat{k}^{b}_{1}\hat{k}^{b'}_{1} f(\bm{k}_{1},\bm{k}_{2}) X_{\rm T}(k_{1}) Y_{\rm T}(k_{2}) 
+ \frac{1}{2}\left( \hat{k}^{b}_{1}\hat{k}^{b'}_{2} + \hat{k}^{b}_{2}\hat{k}^{b'}_{1}\right)
f(\bm{k}_{2},\bm{k}_{1}) X_{\rm T}(k_{2})Y_{\rm T}(k_{1})  \Bigr] \notag \\
&=& 0 ~,
\end{eqnarray}
where we symmetrize about $b\leftrightarrow b'$ in the square bracket by using the nature of the symmetry under the exchange of $k_{1}$ and $k_{2}$.
In conclusion, the Harrison mechanism cannot induce helical magnetic fields.
This result is coming from the fact that general relativity and the standard Maxwell theory do not violate the parity symmetry.
Therefore, the observed helical magnetic fields call for other mechanisms to explain.
\section{Conclusion and Summary}
In this paper, we reinvestigate the spectrum of magnetic fields
induced by cosmological perturbations through the Harrison
mechanism.
If we consider the cosmological perturbation theory up to the first order, the Harrison mechanism does not work since the vector mode, which is needed for this mechanism, has only a decaying solution.
However, when we expand the cosmological perturbations up to the
second order, the regular solution of the vector mode is excited by the
first-order scalar mode. 
The Harrison mechanism works in the higher-order cosmological
perturbation theory. 

In previous studies, the spectrum of magnetic
fields induced by this mechanism has been estimated.
In Ref.~\cite{Ichiki:2007hu}, the authors show the spectrum of second-order magnetic fields induced by the product of the first-order perturbations, namely, the slip and anisotropic stress terms.
Subsequently, in Ref.~\cite{2011MNRAS.414.2354F}, the purely second-order slip term is included.
By comparing these works, however, it is found that 
there are some discrepancies in the product of first-order perturbations.
For example, the power law tails of the spectrum induced by the slip and anisotropic stress terms on large scales have different
$k$-dependences in Refs.~\cite{Ichiki:2007hu} and
\cite{2011MNRAS.414.2354F}.
Furthermore, the scale-dependences of the spectrum are slightly different from each other.
We find that the discrepancy can be
explained by the lack of sampling in the Fourier modes at $k_{1}\approx k_{2}$ of the first-order scaler perturbations in Ref.~\cite{2011MNRAS.414.2354F},
and our results agree with the ones of Ref.~\cite{Ichiki:2007hu}.

Let us summarize features of the magnetic fields induced by the second-order
magnetic fields at cosmological recombination as follows.
\begin{itemize}
\item
The scale dependence of magnetic fields on large scales
 is $\propto k^{3.5}$, which is consistent with the result in
 Ref.~\cite{Ichiki:2007hu}. 
Note that magnetic fields generated by causal processes
have the same power \cite{Durrer:2003ja}. 
\item
On small scales, the spectra of magnetic fields induced by the slip, the
 anisotropic stress, and the second-order slip terms have the power of
 $k^{0.2}$, $k^{1.0}$, and $k^{1.0}$, respectively.
 In particular, the spectra of magnetic fields induced by the slip and anisotropic stress terms are consistent with the result in
 Ref.~\cite{Ichiki:2007hu}.
\item
The cancellation occurs between the anisotropic
 stress term and the second-order slip term on small scales in
 the tight coupling regime in the radiation dominated era,
and the power of magnetic field spectrum becomes $\propto k^{0.2}$.
This result indicates that the spectrum of magnetic fields cannot have
 the large amplitude as argued in Ref.~\cite{Ichiki:2007hu}.
\item
The spectrum of magnetic fields at cosmological recombination has a bump
 at $k\approx 5.0 \times 10^{-1}~h{\rm Mpc}^{-1}$ owing to
 extra amplification after the horizon crossing, where the amplitude of magnetic fields
 is $B_{\rm rec}\approx 5.0\times 10^{-24}~{\rm Gauss}$.
 However, after all, this amplification vanishes by nontrivial flipping of the relative velocity between photons and baryons discussed in Ref.~\cite{2012PhRvD..85d3009I}.
 This cancellation creates characteristic diffusion scales in the magnetic fields spectrum around $k\approx 10^{0}~h{\rm Mpc}^{-1}$.
\item
 The Harrison mechanism does not work efficiency
 below the scale of Silk damping at the electron-positron pair creation epoch as $k_{\rm cut}\approx
 10^{9}~h{\rm Mpc}^{-1}$,
 as discussed in Ref.~\cite{Ichiki:2007hu}. Even if we extrapolate our numerical result
 toward smaller scales by using the power of $\propto k^{0.2}$ up to
 the cutoff scale, 
 the amplitude of the magnetic fields at that
 scale cannot be larger than $10^{-23}~{\rm Gauss}$, assuming that
 the linear density perturbations are scale invariant.
\end{itemize}

Finally, we discuss implications of the cosmological seed fields.
The derived amplitude of magnetic fields at recombination has a peak
about $5.0\times 10^{-24}~ {\rm Gauss}$, which is sufficient for a candidate of the seed of galactic magnetic fields \cite{Davis:1999bt}.
However, this amplitude seems to be somewhat small to explain the intergalactic magnetic fields \cite{2012ApJ...744L...7T,Takahashi:2013lba}.
Note that, the above amplitude is derived assuming that the primordial perturbations are scale-invariant,
while primordial perturbations with a blue tilt lead a larger amplitude of the magnetic fields on smaller scales.

The magnetic fields induced by the second-order perturbation must be inevitably generated in
the standard cosmology,
and it is possibly that the magnetic fields act as seed fields for the turbulent dynamo during
the structure formation of the universe.
\begin{acknowledgments}
We thank Kenji Hasegawa for his comments to understand physical interpretation of the second-order collision term.
This work was supported in part by a Grant-in-Aid for JSPS Research under Grant No.~26-63 (S.S.) and a Grant-in-Aid for JSPS Grant-in-Aid for Scientific Research under Grants No.~24340048 (K.I.) and 25287057 (N.S.).
We also acknowledge the Kobayashi-Maskawa Institute for the Origin of Particles and the Universe, Nagoya University, for providing computing resources useful in conducting the research reported in this paper.
\end{acknowledgments}
\bibliography{ref}
\end{document}